\documentclass[aps,pre, onecolumn, 10pt]{revtex4-2}
\usepackage{amsmath}
\usepackage{graphicx}
\usepackage{xcolor}
\usepackage[applemac, utf8]{inputenc}
\usepackage[french]{babel}
\usepackage[T1]{fontenc} 
\usepackage{amssymb}
\usepackage{ulem} 
\usepackage{soul} 
\usepackage{bm}

\usepackage{hyperref} 

\begin{document}

\title{Hydrodynamics and propulsive efficiency of stern-sculling propulsion}
\author{G. Roudaut$^{1}$}
\email{contact@universitedelagodille.org}
\author{B. Augier$^2$}
\author{M. Rabaud$^3$}
\affiliation{$1$ Association Tud Yaouank Ar Mor, 29810 Lampaul-Plouarzel, France.\\
$2$ IFREMER, RDT Research and Technological Unit, 1625 Route de Sainte-Anne, 29280, Plouzan\'e, France.\\
$3$ Universit\'e Paris-Saclay, CNRS, FAST, F-91405 Orsay, France. \\}

\date{\today}

\begin{abstract}

Stern-sculling is a traditional form of human-powered boat propulsion in which a single oar is oscillated laterally at the stern. In this study, we investigate this propulsion technique from a hydrodynamic perspective by continuously measuring the oar kinematics, the boat motion, and the forces acting on the oar across a broad range of boat speeds. During the lateral strokes, lift serves as the main propulsive force, whereas during the backward strokes, propulsion is primarily associated with added-mass effects. These measurements allow us to determine the propulsive efficiency as a function of boat speed. The efficiency peaks at approximately 65\% at an intermediate speed and then decreases progressively to approximately 25\% at higher speeds. These results provide insight into the hydrodynamic mechanisms governing this traditional mode of propulsion and highlight the effectiveness of stern-sculling at moderate speeds.

\end{abstract}

\maketitle

\section{Introduction} \label{sec:intro}

Stern-sculling, also known as single-oar sculling, is a traditional propulsion technique in which a single oar remains in contact with the transom while its blade stays immersed in the water. Although this form of human-powered propulsion is still practiced in several parts of the world, it has been studied primarily from ethnological and biomechanical perspectives~\cite{Roudaut_2015, joulanda2015experimental, takayamameasurement}. Here, we aim to characterize stern-sculling from a fluid-mechanical perspective.

\begin{figure}[htb]
\begin{center}
\includegraphics[width=0.35\textwidth]{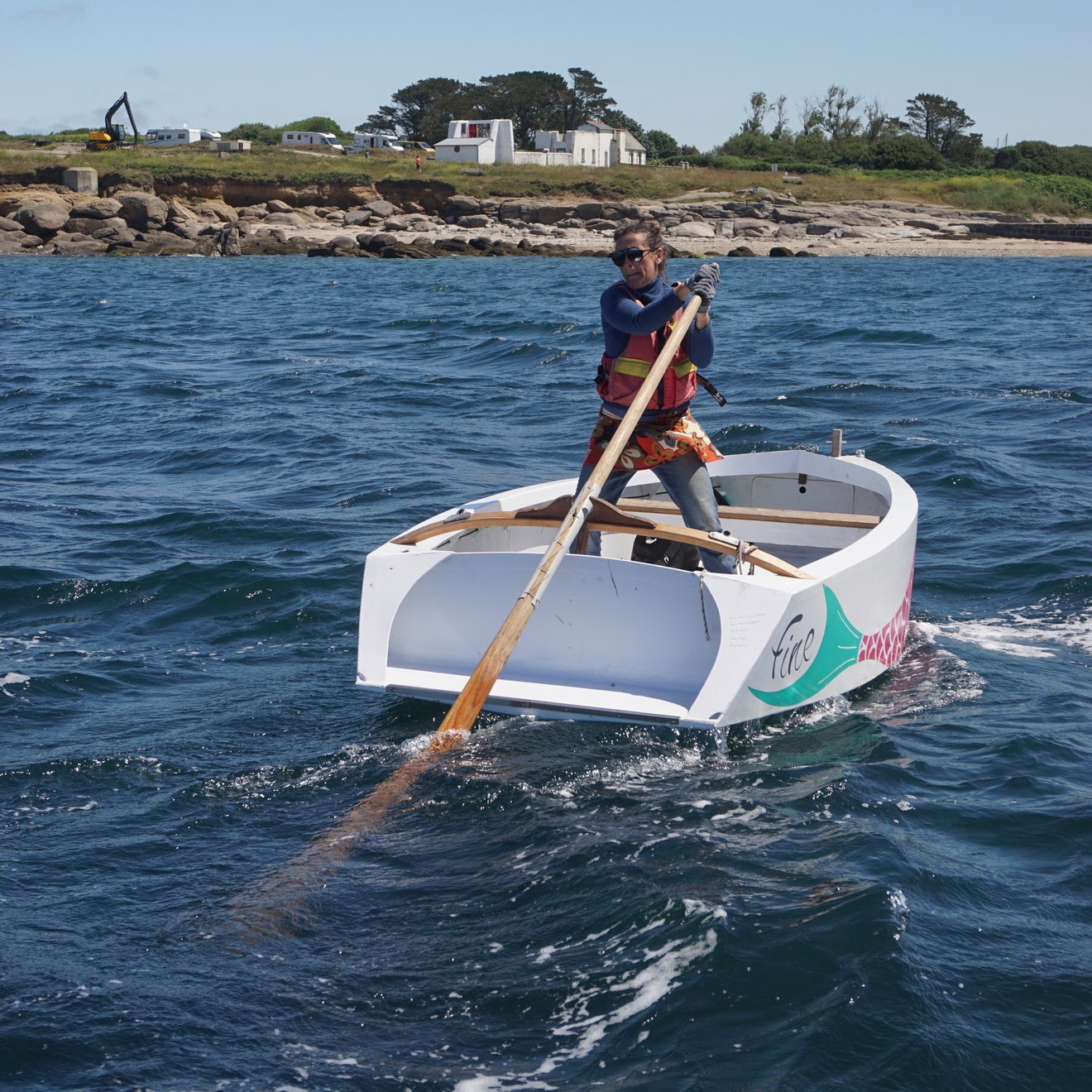}
\includegraphics[width=0.526\textwidth]{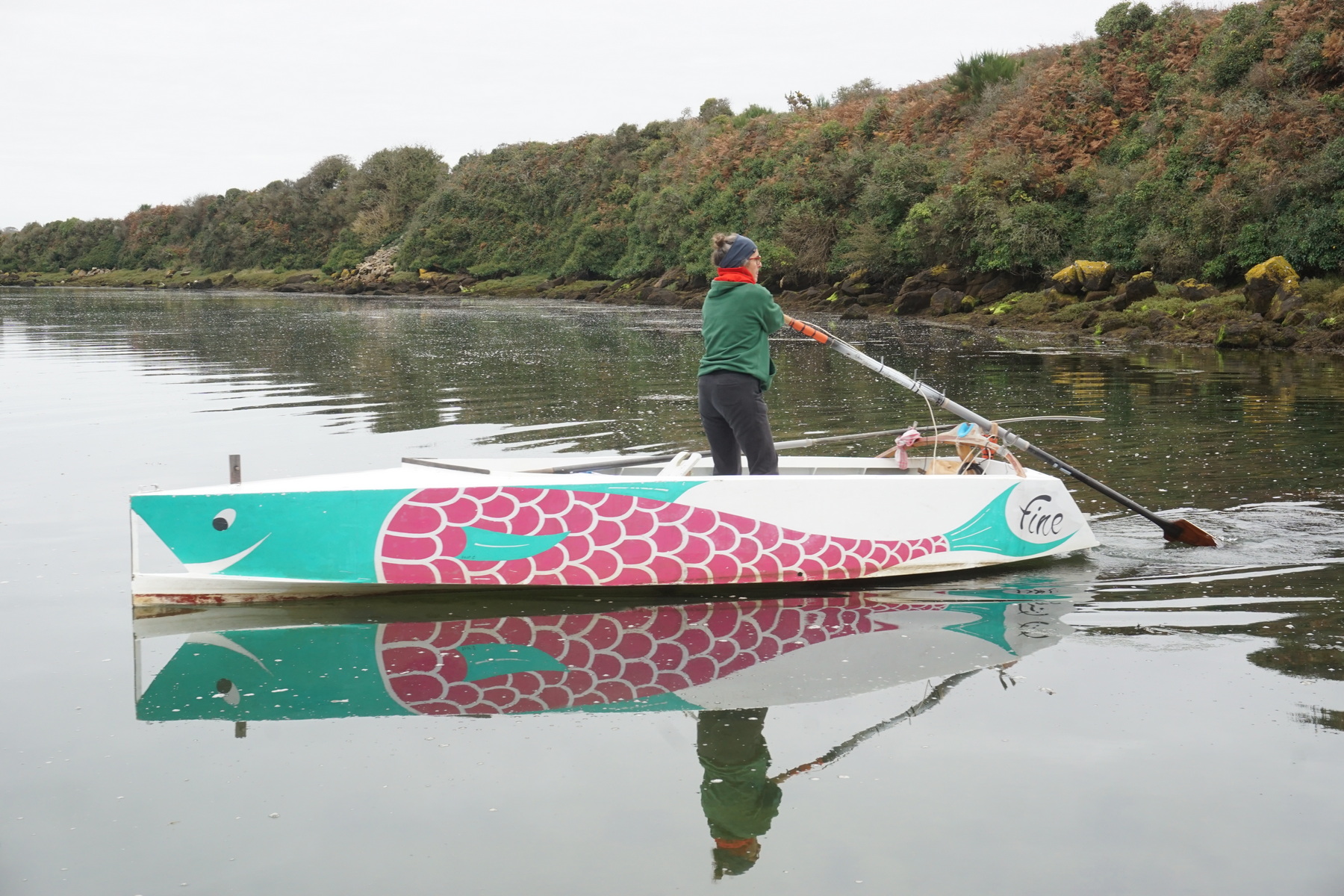}
\caption{Photographs of two scullers performing stern-sculling, facing backward, aboard the boat used in this study: (left) rear view; (right) port-side view. The oar rests in an oarlock mounted on the transom. During sculling, the hands execute a horizontal back-and-forth motion similar to the infinity symbol "$\infty$". }
\label{Photo_godilleur}
\end{center}
\end{figure}

The sculler, who usually faces backward, holds the oar at an angle to the water surface (Fig.~\ref{Photo_godilleur}). The hands oscillate the upper part of the oar laterally, while an intermediate point of the oar is seated in a notch or oarlock mounted on the transom. The hands follow a trajectory resembling the infinity symbol "$\infty$", while simultaneously rotating the oar about its axis to adjust the angle of attack. The immersed part of the oar is wider than the part above the water and forms a blade that remains submerged at all times. By adjusting the blade's orientation during each lateral stroke and subsequent reversal, the sculler can generate propulsive forces throughout all phases of the cycle.

For this propulsion technique to be effective, it is convenient to use a rather heavy boat with a long keel. Indeed, because the oar applies alternating lateral forces to the stern, a lightweight or keel-less boat, such as an inflatable dinghy, can undergo substantial yaw oscillations that significantly reduce the propulsive force. Stern-sculling nevertheless offers several practical advantages: it requires only one oar, the boat is easy to steer, moves naturally straight, it can be used even on a large boat, and it does not take up much space, which is an advantage in a crowded anchorage. Once learned, the motion is relatively continuous and can drive the boat at speed of the order of $1\,\text{m/s}$ with high power efficiency. However, maintaining the stroke becomes more difficult and propulsion becomes less effective at higher speeds.

Kayaking \cite{hemon2018hydrodynamic}, canoeing, and rowing (see \cite{grift2021hydrodynamics} and references therein) rely primarily on drag forces on the oars and include a recovery phase with the blade out of the water. In contrast, stern-sculling also uses lift forces generated by the blade's interaction with the water to propel the boat forward.
During the lateral strokes the blade's edges alternately serve as the leading and trailing edges, requiring a blade profile that is left/right symmetrical. The blade generates lift during these lateral strokes, while the side facing the surface remains the pressure side. The position of the blade at the end of these lateral movements is perpendicular to the water flow. This propulsion technique allows the use of cambered, yet symmetrical, blade profiles.

\begin{figure}[htb]
\begin{center}
\includegraphics[width=0.78\textwidth]{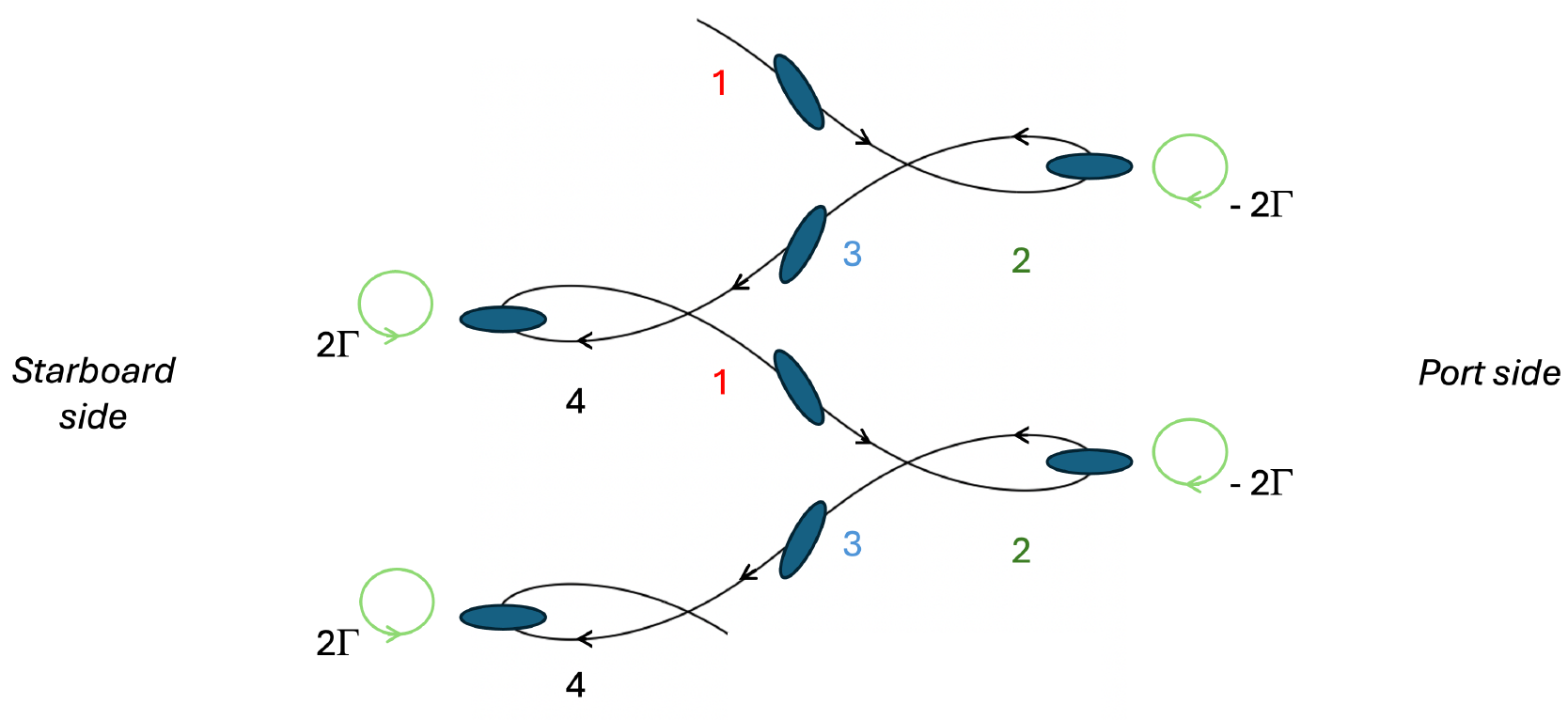}
\caption{Schematic of the oar blade trajectory in the water reference frame during two periods. The oar's orientations are shown in each phase, along with the positions of the vortices released during phases 2 and 4 (green).}
\label{Fig_Lemni}
\end{center}
\end{figure}

Figure~\ref{Fig_Lemni} illustrates the trajectory and orientation in the water frame of the oar's blade, with the boat moving downward. The stroke cycle is divided into four phases; during phase 1 the blade moves toward port; during phase 2 the sculler rotates the oar counterclockwise about its axis; during phase 3, the blade moves toward starboard; and during phase 4, the oar is rotated clockwise around its axis. As discussed below, lift is generated during phase 1 and 3, giving rise to opposite circulations $\pm \, \Gamma$ of the velocity around the blade. During phase 2 and 4 the corresponding changes in circulation lead to the shedding of twice of this circulation as start and stop vortices \cite{guyon2015physical, pullin1980some}. These vortices, when released, generate a Bénard-von K\'arm\'an vortex street (a staggered row of vortices) corresponding to a net flux of water moving backwards as the boat is pushed forward by the oar.
 
This mode of operation is unique compared to other propulsion technique used in the animal kingdom. Although figure-eight trajectories occur in several biological systems, the pectoral and caudal fins of fish, as well as the wings of insects and birds, generally retain a consistent leading edge during a stroke~\cite{farisenkov2022novel}. Related hand motions have nevertheless been described as "water-treading" in front-crawl swimming~\cite{takagi2014unsteady} and also been explored in some micro-air-vehicle projects~\cite{ lua2017water,krishna2022tread}.

For the present study, we instrumented a boat to measure the oar's kinematics and the forces acting on it. These measurements were used to determine the mean power supply to the oar and the mean hydrodynamic propulsive power, allowing us to quantify the propulsive efficiency over a range of boat speeds.
Two experimental runs were conducted, each with a different experimented sculler. Both runs were performed on calm water, in the absence of appreciable wind or current. Each run began with the boat at rest, after which its speed was gradually increased until the maximum speed attainable by the sculler.

\section{Description of the boat, oar and sensors}
\label{Sec_Setup} 

\subsection{Data on the boat and the oar}

The boat used in this study is shown in Figure~\ref{Photo_godilleur}. It measures 4.74 m in length (both overall and at waterline), 1.63 m in width, and has a design draft of 0.175 m. Constructed from marine plywood, its unladen weight is approximately 100 kg. It has a hard chine hull with a long and shallow keel (see \url{https://universitedelagodille.org/plans-du-scaff/} for more details). This boat was chosen for its high-speed performance (with a maximum speed of $10\,\text{km/h}$ for experienced scullers), good course-keeping ability, and superior roll stability compared to more conventional sculling boats. The oarlock is centered on the transom and located at 0.6 m above the water surface. We neglect the small lateral motion of the oar within the oarlock and model the oar--boat connection as a ball joint.

\begin{figure}[htb]
 \begin{center}
 \includegraphics[width=1\linewidth]{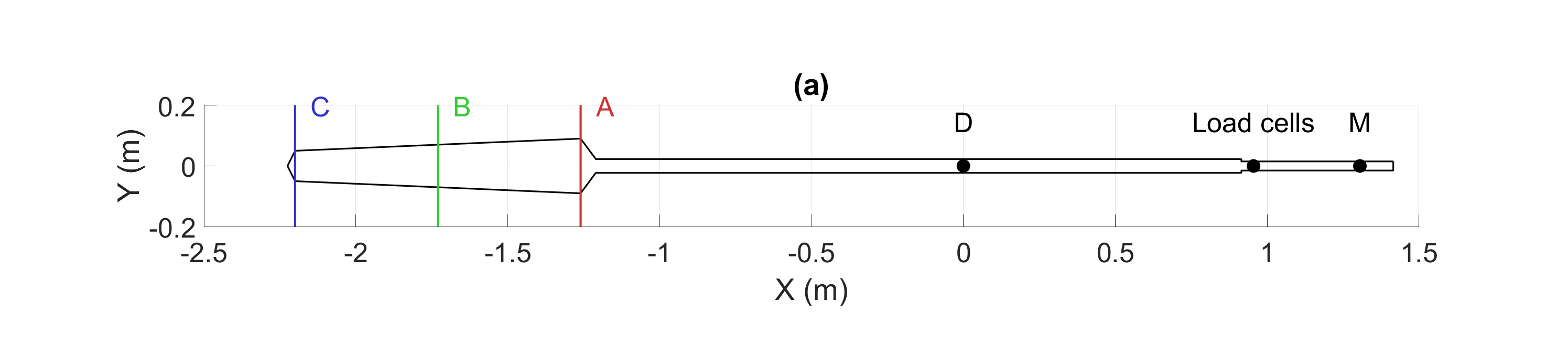}
\includegraphics[width=0.8\textwidth]{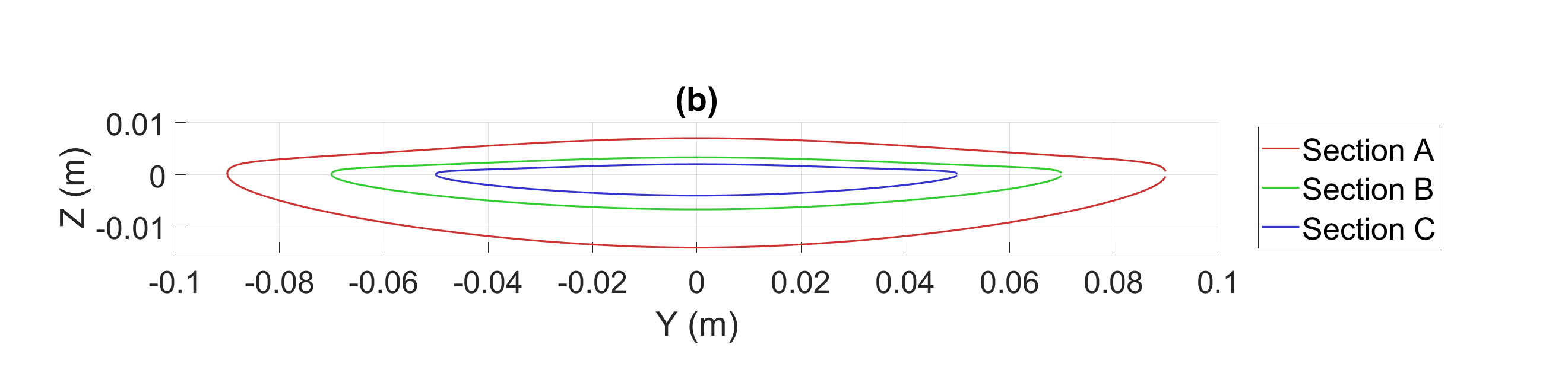}
 \caption{Geometry of the oar used in this study. (a) Frontal view (in the plane $Z=0$) where the load cells location is visible; (b) Three blade sections in planes of constant $X$: 
 A (red, $X_A= - 1.26$ m), B (green, $X_B= - 1.73$ m) and 
 C (blue, $X_C= - 2.20$ m). The sections show that the blade is left---right symmetric and asymmetric with respect to $Z$, with a small upward camber of 2\%. }
 \label{oar_drawing}
\end{center}
\end{figure}

The oar shaft is a straight cylinder made of carbon fiber, with a diameter of 55 mm. The total length of the oar is $L=3.62\,\text{m}$ (Fig.~\ref{oar_drawing}a). The blade is made of wood, measures $1\,\text{m}$ in length, and is $180\,\text{mm}$ wide at the top and $100\,\text{mm}$ at the bottom (Fig.~\ref{oar_drawing}b). The mean wingspan-to-chord ratio is thus $AR =7.1$. The blade was designed to display homothetic sections.
The maximum thickness of the profile is $21\,\text{mm}$ at the top and $6\,\text{mm}$ at the bottom. Figure~\ref{oar_drawing}b shows three sections of the blade. They are left/right symmetric with a relative thickness of 12\%, slightly cambered (2\%), and with a flatter upper side ($Z > 0$) which is always the pressure side. 

Three reference points on the oar are identified: point $M$ where the force exerted by the hands $\bm F_M$ is applied on the oar handle, point $D$ where a force $\bm F_D$ is exerted by the oarlock on the oar, and point $H$, located on the blade, at which the resultant hydrodynamic force $\bm F_H$ is assumed to act. The position of $H$ is not known \textit{a priori} and may vary slightly along the oar during its periodic motion. We assume that $H$ remains on the oar axis, such that $H(X,Y,Z)=(X_H,0,0)$. Note that, the force $F_D$ must remain positive upward throughout the movement when using a notch, so that the oar does not lift out of the notch, which sometimes happens to beginners! 

The oar has a mass of $m=2.92\,\text{kg}$.
When fully out of the water, its center of gravity $G$ is located below point $D$ ($X_G<0$). However, once submerged, buoyancy acting on the submerged portion of the oar makes the net moment about $D$ approximately zero.
The mean angle of immersion of the oar with respect to the horizontal water surface is of the order of $30^\circ$. The distance $DM$ between the oarlock and the hands is $1.30\,\text{m}$.
We choose a rather stiff oar: a maximum of $40\,\text{mm}$ deflection is measured when a force of $80\,\text{N}$ is applied at point $M$ with point $D$ and the blade fixed. We will therefore treat the oar and the blade as rigid bodies.

\subsection{Frames of reference and orientation angles}

Three frames of reference are introduced to describe the motions of the oar and the boat.

\begin{figure}[htb]
\begin{center}
\includegraphics[width=0.9\textwidth]{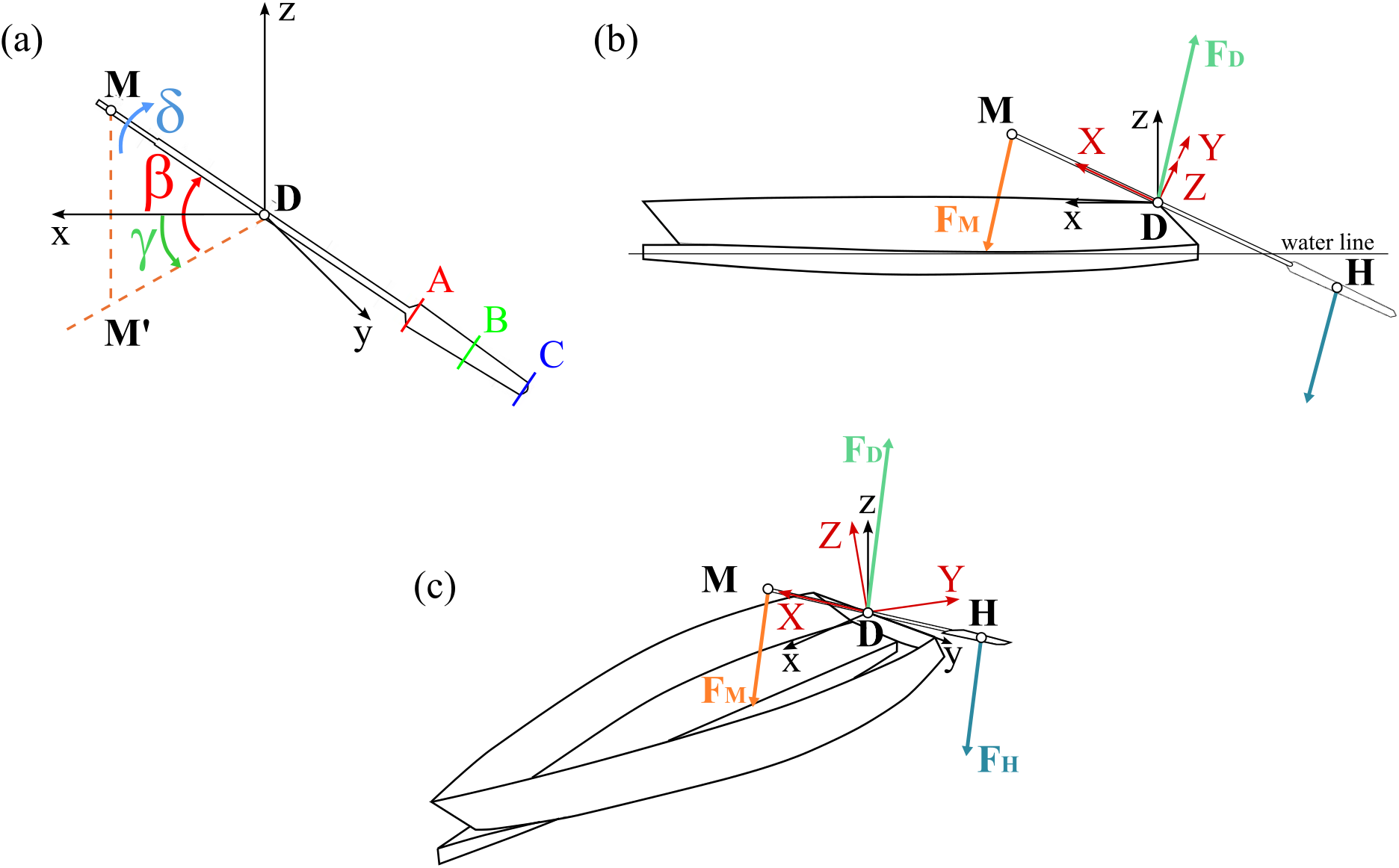}
\caption{Sketch of the oar on the boat illustrating the oar frame $(D,X,Y,Z)$ and the boat frame $(D,x,y,z)$: (a) perspective view of the orientation angles of the oar in the boat's frame: yaw ($\gamma$), pitch ($\beta$) and roll ($\delta$). The three forces $\bm F_M$, $\bm F_D$ and $\bm F_H$ are applied to the oar at points $M$, $D$ and $H$, respectively; (b) side view in the plane $y=0$; and (c) perspective view.}
\label{Fig_referentiels}
\end{center}
\end{figure}

 \begin{itemize}
 \item \textbf{The oar frame} (Fig.~\ref{oar_drawing} and Fig.~\ref{Fig_referentiels}b and c) $(D,\bm e_X, \bm e_Y, \bm e_Z)$, has its $X$ axis along the oar pointing towards the point $M$, its $Z$ axis is perpendicular to the blade plane and has a positive vertical component, and its $Y$ axis is defined so that the frame is right-handed (with $\bm e_Y$ pointing on average towards port).
\item \textbf{The boat frame} (Fig.~\ref{Fig_referentiels}b and c) $(D,\bm e_x, \bm e_y, \bm e_z)$, has its $x$ axis along the boat centerline and pointing forwards, its $y$ axis pointing towards port and lying in the horizontal plane when the boat is at rest, and its $z$ axis pointing upward, so that the frame is right-handed.
\item \textbf{The mean boat frame}. Because the boat undergoes small oscillations during propulsion, we also introduce a reference frame in which the kinematics are evaluated. During the tests, the boat rotates around its instantaneous center of rotation. We assume that this point, denoted by $C_g$, remains fixed in the boat frame and coincides with the centroid of the submerged hull cross section in the plane $y=0$. Its coordinates in the boat frame are $C_g(x, y, z) = (1.9\,\text{m}, 0, -0.63\,\text{m})$.
This mean boat frame is defined as $(C_g,\bm e_{x'}, \bm e_{y'}, \bm e_{z'})$, where $x'$ is aligned with the mean boat trajectory and points forward, $y'$ lies in the horizontal plane, perpendicular to $x'$ and points on average towards port and $z'$ points vertically upward.
 \end{itemize}

Three angles are also necessary to describe the orientation of the oar in the boat frame (Fig.~\ref{Fig_referentiels}a): $\gamma$ is the oar yaw angle, defined as the angle between the boat's $x$ axis and the projection of the oar's $X$ axis onto the $(x,y)$ plane, with positive values towards port; $\beta$ is the oar pitch angle, defined as the angle between the projection of $X$ axis in the plane $(x,y)$ and the $X$ axis, with positive values upward; and $\delta$ is the oar roll angle, defined as the rotation angle of the oar about its $X$ axis, with positive value counterclockwise. The reference configuration $\delta=0$ is defined by $\bm e_Z$ lying in the $(x,z)$ plane.

\subsection{Sensors and acquisition protocol}

\begin{figure}[htb]
\begin{center}
\includegraphics[width=0.7\textwidth]{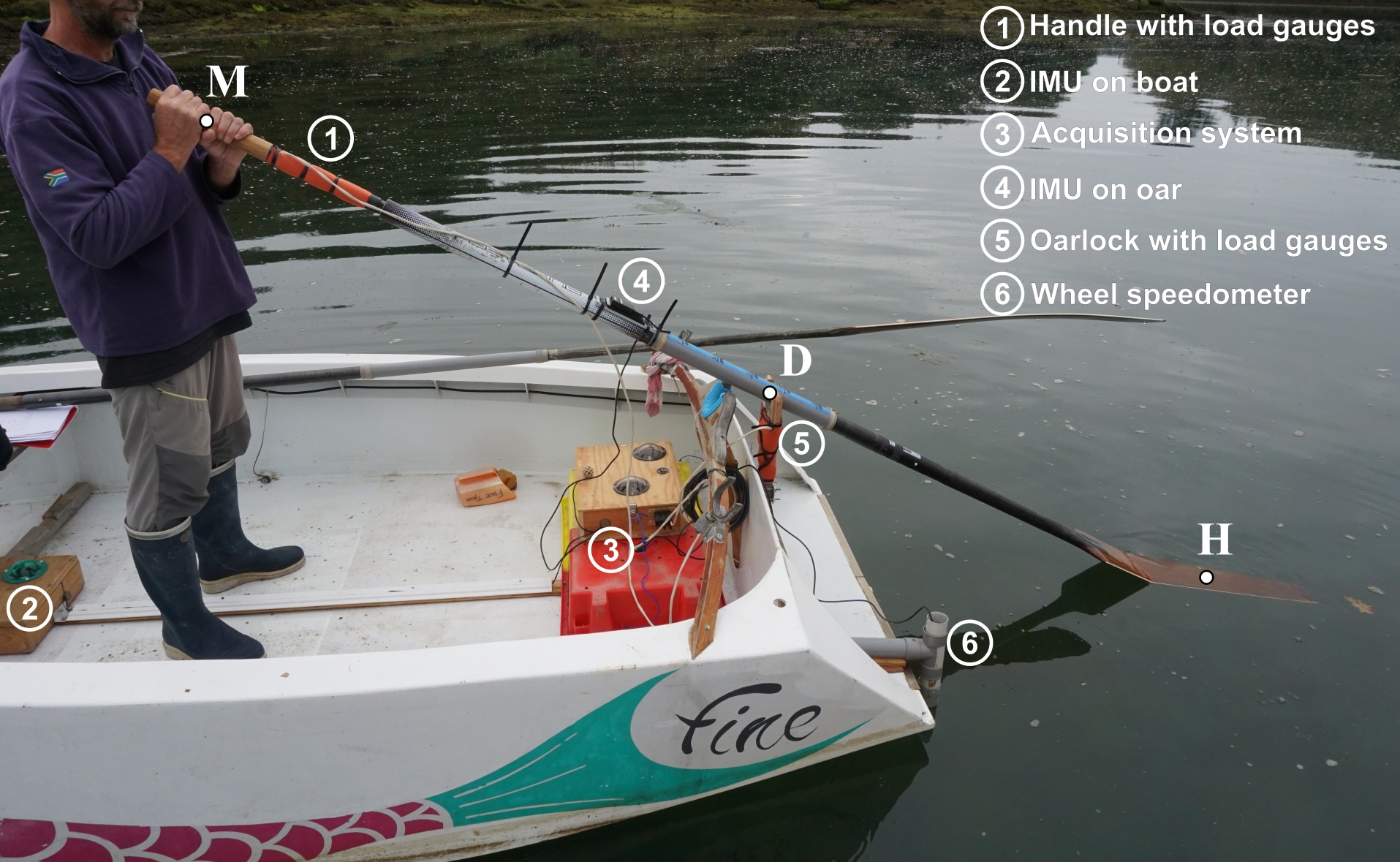}
\caption{View of the acquisition system and sensors mounted at the stern of the boat.} \label{Photo_sensors}
\end{center}
\end{figure}

To measure the kinematics and the forces, the boat is equipped with various sensors (Fig.~\ref{Photo_sensors}):

\begin{itemize}
 \item At the oarlock, point $D$, the two components $F_{Dx}$ and $F_{Dy}$ of the force $\bm F_D$ are measured in the boat's frame with resistive gauges.
 \item At point $M$ the two components $F_{MY}$ and $F_{MZ}$ of the force $\bm F_M$ are measured in the oar frame by the instrumented handle.
 \item Two inertial measurement units (IMU) are mounted on the oar and inside the boat. The oar-mounted IMU (a \textit{3space USB sensor} from Yost Labs) is located close to the oarlock, whereas the boat-mounted IMU (a \textit{Bricklet 3.0} from Tinkerforge) is fixed close to the center of gyration of the hull. 
 \item The boat's speed relative to the water is measured by a paddle wheel speedometer calibrated using GPS positions.
\end{itemize}
 All sensors are connected to a single-board computer (a Raspberry Pi 3B+) and the data are recorded in real time. The sampling rate is $50\,\text{Hz}$ for all sensors, and the data streams are synchronized. A GPS unit (GP102 model from Canmore) is also installed on board. It provides the boat's speed over ground at $1\,\text{s}$ intervals, based on GPS signal analysis.

The data provided by the IMUs are used to compute all the rotations, enabling all measured quantities to be expressed in any of the three reference frames.
The boat IMU provides the instantaneous orientations and angular velocities of the hull relative to the mean boat frame. The measured yaw, roll and pitch oscillations of the hull remain small at moderate velocity, reaching $\pm 4 ^\circ$, $\pm 3 ^\circ$, and $\pm 1 ^\circ$, respectively, at the highest velocities ($U=2.2\,\text{m/s}$). From these measurements we determine the position of the oarlock $D$ in the mean boat frame at all times.
The oar IMU gives the instantaneous orientations and angular velocities of the oar in the mean boat frame. The oar moves with the boat and also rotates around $D$. Finally, the measured position of $D$ and the orientation of the oar are used to reconstruct the positions of the blade and point $M$ in the mean boat frame.

\subsection{Forces equilibrium}

Newton's second law applied to the oar writes:

\begin{equation}
\bm F_M + \bm F_D + \bm F_H =m \bm a_G,
\label{eqn.forces}
\end{equation}
where $\bm F_M$, $\bm F_D$ and $\bm F_H$ are the three forces  previously defined, $m$ is the mass of the oar, and $\bm a_G$ is the acceleration of oar's center of mass $G$ relatively to an inertial frame. 

However, this inertial force term is small because the oar has a relatively low mass, and, when the oar is immersed, its center of mass $G$ lies close to $D$, which undergoes only small acceleration. Our experimental data indicate that this inertia term is less than a few percent of the applied forces. 
Thus, we assume that the sum of the three forces vanishes, implying that the three forces are coplanar.
There is usually no component of $\bm F_M$ along the oar axis ($F_{MX}=0$) because a skilled sculler can move the oar by applying a force normal to its axis. Similarly, the force exerted by the oarlock at $D$ has practically no component along $X$ ($F_{DX}=0$), because no sliding is observed at $D$ even though this movement would be possible within the notch due to the low friction (PVC on PVC). Since the sum of the three forces is assumed to vanish, the same holds for $\bm F_H$: $F_{HX}\approx0$. Consequently, the three forces are lie approximately to the plane $(Y,Z)$, perpendicular to the oar.


The gauges on the handle measure the two components of the force exerted by the sculler on the oar, $F_{MY}$ and $F_{MZ}$. With $F_{MX}=0$, the force $\bm F_M$ is therefore fully determined.
The gauges on the oarlock measure only the two horizontal components, $F_{Dx}$ and $F_{Dy}$, of the force exerted by the oarlock on the oar. The third component, $F_{Dz}$, is not measured but can be calculated by imposing that the force at $D$ has no component along the oar axis $X$.
Indeed, as illustrated in Fig.~\ref{Fig_referentiels}c, we have $$F_{DX}=\cos \gamma\cos \beta F_{Dx}+\sin \gamma\cos \beta F_{Dy}+\sin \beta F_{Dz} = 0,$$ which gives: 

\begin{equation}
F_{Dz}=-\frac{\cos \gamma}{\tan \beta} F_{Dx} -\frac {\sin \gamma}{\tan \beta}F_{Dy}.
\label{eqn.FDz}
\end{equation}

Thus, $\bm F_D$ is also fully known and the hydrodynamic force $\bm F_H$ can then be computed at all times. We are particularly interested in its component along $x'$ in the mean reference frame of the boat, $F_{Hx'}$, and in its time average, $\overline{F_{Hx'}}$, which represents the mean propulsive force acting on the boat.

\section{Temporal evolution of the propulsive force and of the injected power}

\begin{figure}[htb]
\begin{center}
\includegraphics[width=0.9\textwidth]{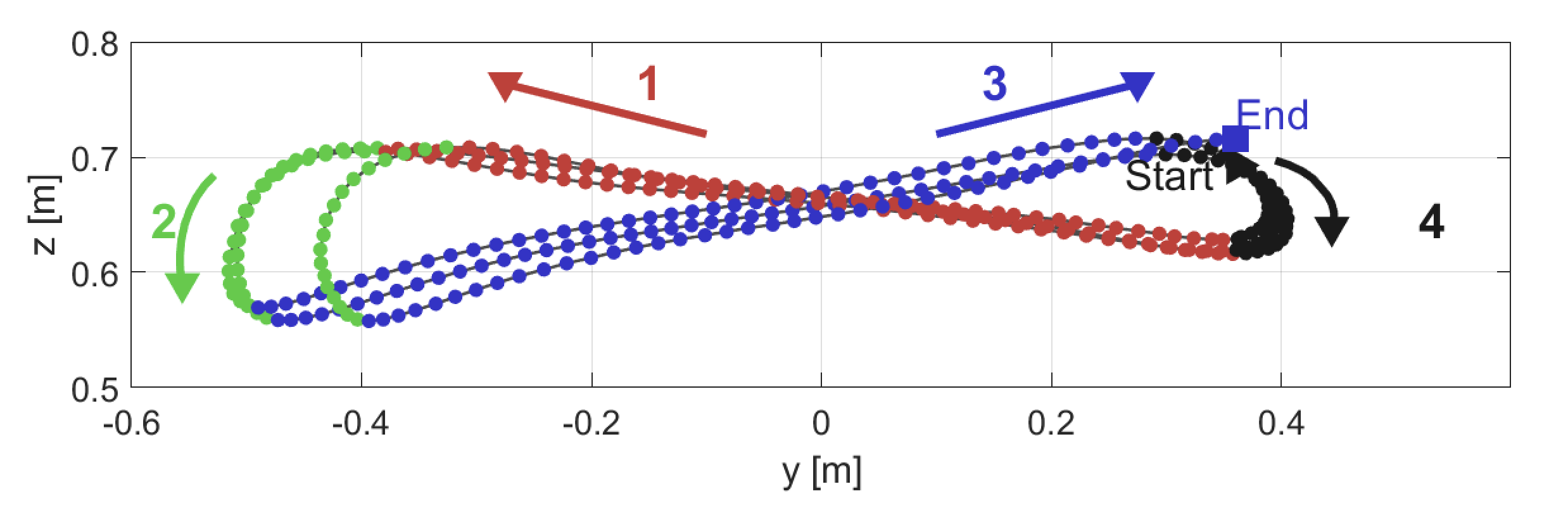}
\caption{Displacement of the oar handle relative to the boat during three consecutive cycles, for the boat moving at $U = 1.3$ ~\text{m/s}. The center of the handle actually moves on a sphere, but its projection in the vertical plane is plotted here (looking backward). Time interval between points is $201\,\text{ms}$ and mean period of the "$\infty$" motion is $2.31\,\text{s}$. Numbers and colors illustrate the four phases defined in the text.}
\label{Fig_Mvt_8}
\end{center}
\end{figure}

 The measured displacements of the oar handle are presented in Figure~\ref{Fig_Mvt_8}. The four phases discussed in Figure~\ref{Fig_Lemni} are identified by color: phase 1 (red) where the hands move upward to starboard; phase 2 (green) where the wrists rotate counterclockwise; phase 3 (blue) where the hands move upward to port side; and phase 4 (black) where the wrists rotate clockwise. At this instant, the peak-to-peak lateral displacement of the hands is $A_M \approx 0.91\,\text{m}$, and the small left/right asymmetry visible on this plot is due to a slight asymmetry of the sculler gesture.

\begin{figure}[htb]
\begin{center}
\includegraphics[width=0.8\textwidth]{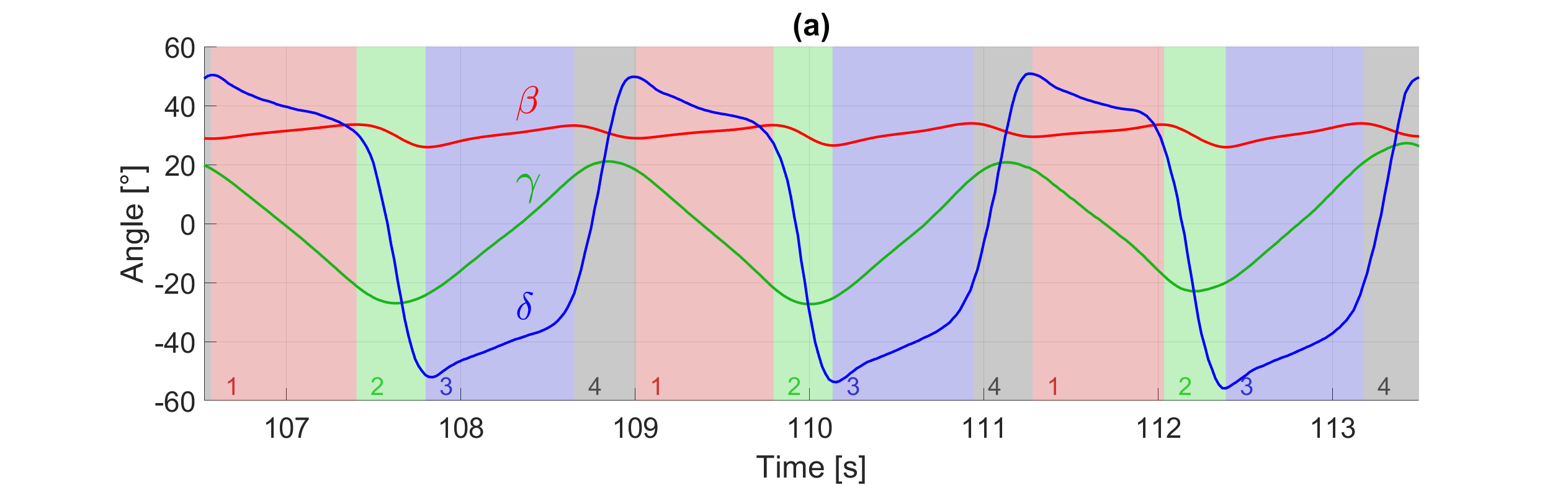}
\includegraphics[width=0.8\textwidth]{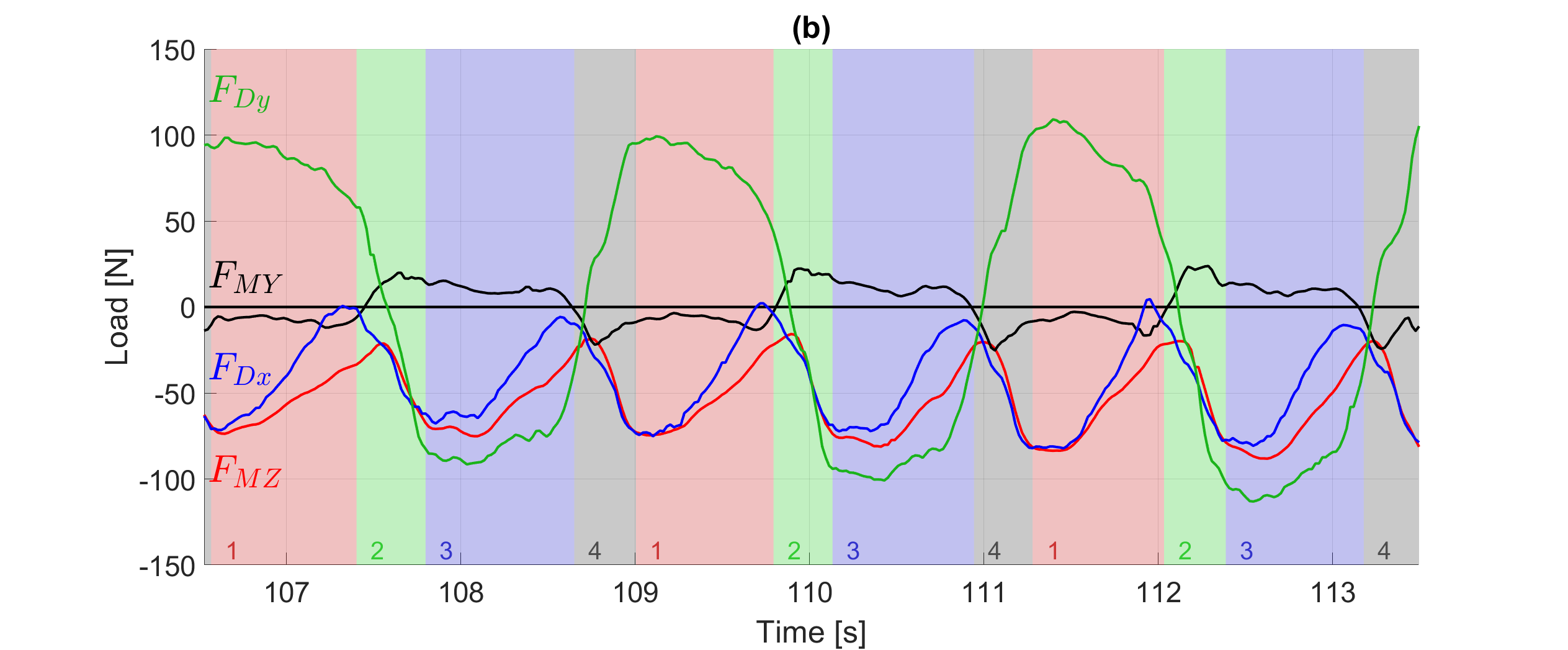}
\includegraphics[width=0.8\textwidth]{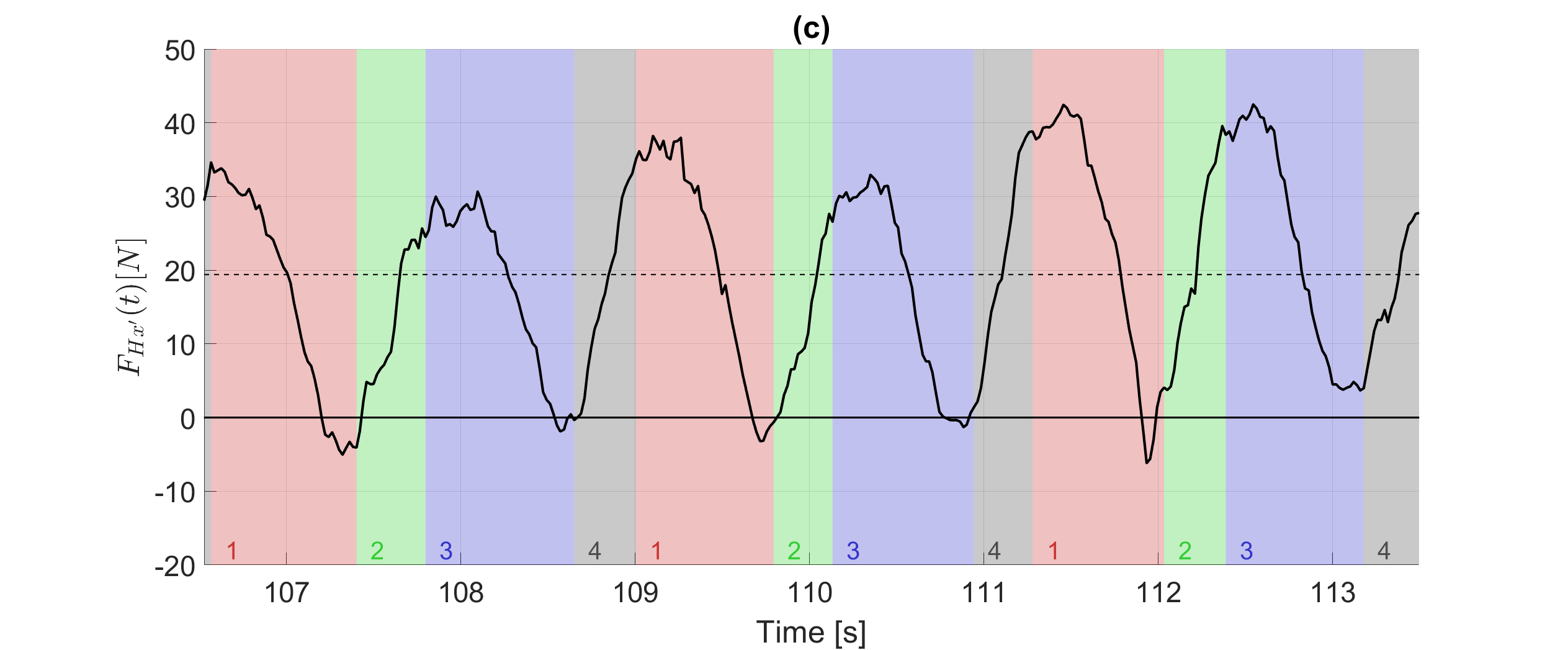}

\caption{Temporal evolution of oar angles and forces measured during three cycles at boat speed $U=1.2$ ~\text{m/s}. Colored areas and numbers above the abscissa indicate the four phases previously described in Fig~\ref{Fig_Mvt_8}: 
(a) Oar orientation angles measured in the boat's frame ($x,y,z$): pitch $\beta$ (red line), yaw $\gamma$ (green line) and roll $\delta$ (blue line). 
(b) Measured components $F_{Dx}$ (blue line) and $F_{Dy}$ (green line), exerted by the oarlock and $F_{MY}$ (black line) and $F_{MZ}$ (red line) exerted on the handle. 
(c) Propulsive force $F_{Hx'}$ (in the direction of the mean trajectory), deduced from Eq.~\ref{eqn.forces}. The black dashed line represents the mean propulsive force during the 3 cycles: $\overline{F_{Hx'}} = 19.41\,\text{N}$.}

\label{3_Evolutions}
\end{center}
\end{figure}

Figure~\ref{3_Evolutions}a illustrates the evolution of the oar orientation angles during three successive periods of propulsion. The results indicate that the pitch angle oscillates only slightly around $\beta \approx 30 ^\circ$, while the yaw angle oscillates in a nearly triangular waveform between + 20$ ^\circ$ and - 20$ ^\circ$. The roll angle behaves more like a square wave, with a slow decrease during the lateral phase (1 and 3) and rapid inversion during phases 2 and 4.

Figure~\ref{3_Evolutions}b shows the temporal evolution of the four components of the forces measured by the resistive gauges. The component $F_{Dx}$ remains negative throughout, reaching extrema at the beginning of phases 1 and 3 and approaching zero at the start of phases 2 and 4. This small negative value indicates that the oar is close to losing contact with the oarlock at this moment. The component $F_{Dy}$ reaches its maximum during the lateral phases 1 and 3. The component $F_{MZ}$ is always negative and reaches its maximum during the lateral phases, while $F_{MY}$ remains always small as the blade is quite thin. 

 Figure~\ref{3_Evolutions}c shows the propulsive component $F_{Hx'}$. It is positive, except for short periods at the end of phase 1 and 3 when the roll angle is smaller, and reaches a maximum at the beginning of these lateral phases 1 and 3. Nevertheless, all four of these phases are propulsive. As in Figure~\ref{Fig_Mvt_8} the slight left/right asymmetry in sculler A's movement is visible, but the three strokes are fairly consistent. Such asymmetry is not unusual and is probably related to the fact that scullers are left or right-handed, and/or to the fact that one hand is higher than the other on the oar.

\begin{figure}[htb]
\begin{center}
\includegraphics[width=0.7\textwidth]{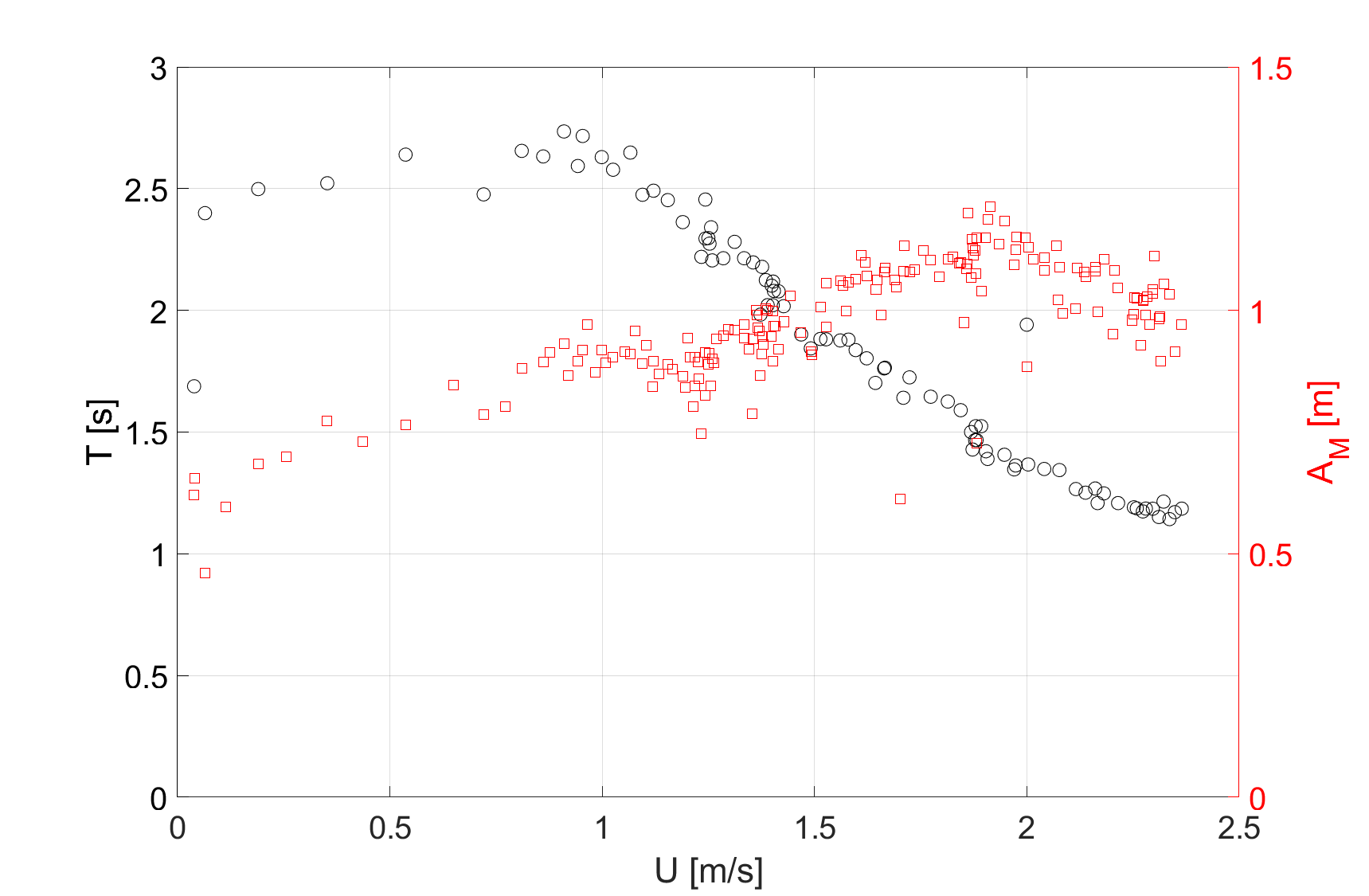}
\caption{Evolution of the period $T$ ($\scalebox{0.7}{\(\bigcirc\)}$) and peak-to-peak amplitude $A_M$ (\textcolor{red}{$\square$}, measured along $y$ in the boat's frame) of the hand excursion of sculler $A$ versus the boat speed.}
\label{Fig_Ampl_period}
\end{center}
\end{figure}

To investigate the dependence of the measured quantities on the boat speed, we conducted two experimental runs in calm water, with no current and wind speed below 2 knots. Each run was performed by a different sculler ($A$ or $B$) and consisted of accelerating from rest to the maximum speed attainable by that sculler. The boat acceleration remained small (below $0.02\,\text{m/s}^2$) so we will consider that the boat is in quasi-static condition at all speeds.
Figure~\ref{Fig_Ampl_period} shows the evolution of the period $T$ and the peak-to-peak lateral displacement of the hands $A_M$ for sculler A during the corresponding run. Similar trends are observed for sculler B. At low velocity the period remains approximately $2.5\,\text{s}$ until $1 \,\text{m/s}$ and then decreases almost linearly at larger velocity. Meanwhile, the amplitude of the hand movements increases until the boat speed reaches $2\,\text{m/s}$, and then decreases slightly. 
To produce a given mechanical power, a sculler can make wide and slow movements of the hands or small and quick ones. The choice of a given amplitude and period $(T,A_M)$ usually corresponds to a minimization of the human effort (optimal muscular efficiency), as demonstrated in many sport such as running or cycling \cite{dickinson1929efficiency, hintzy1999optimal} or in the gaits of animals or humans \cite{hogberg_1952, Alexander_1989}.

From the data of Figure \ref{Fig_Ampl_period}, we can estimate the order of magnitude of the blade's lateral velocity. For example, at the level of the section $B$ of the blade, this typical speed is given by:

\begin{equation}
u_B=\frac{|X_B|}{X_M}\frac{A_M}{T}.
\label{eqn_u_B}
\end{equation}

This velocity can be made dimensionless using the boat speed, yielding the Strouhal number:

\begin{equation}
St=\frac{u_B}{U}.
\label{eqn_Strouhal}
\end{equation}

\begin{figure}[htb] 
\begin{center}
\includegraphics[width=0.7\textwidth]{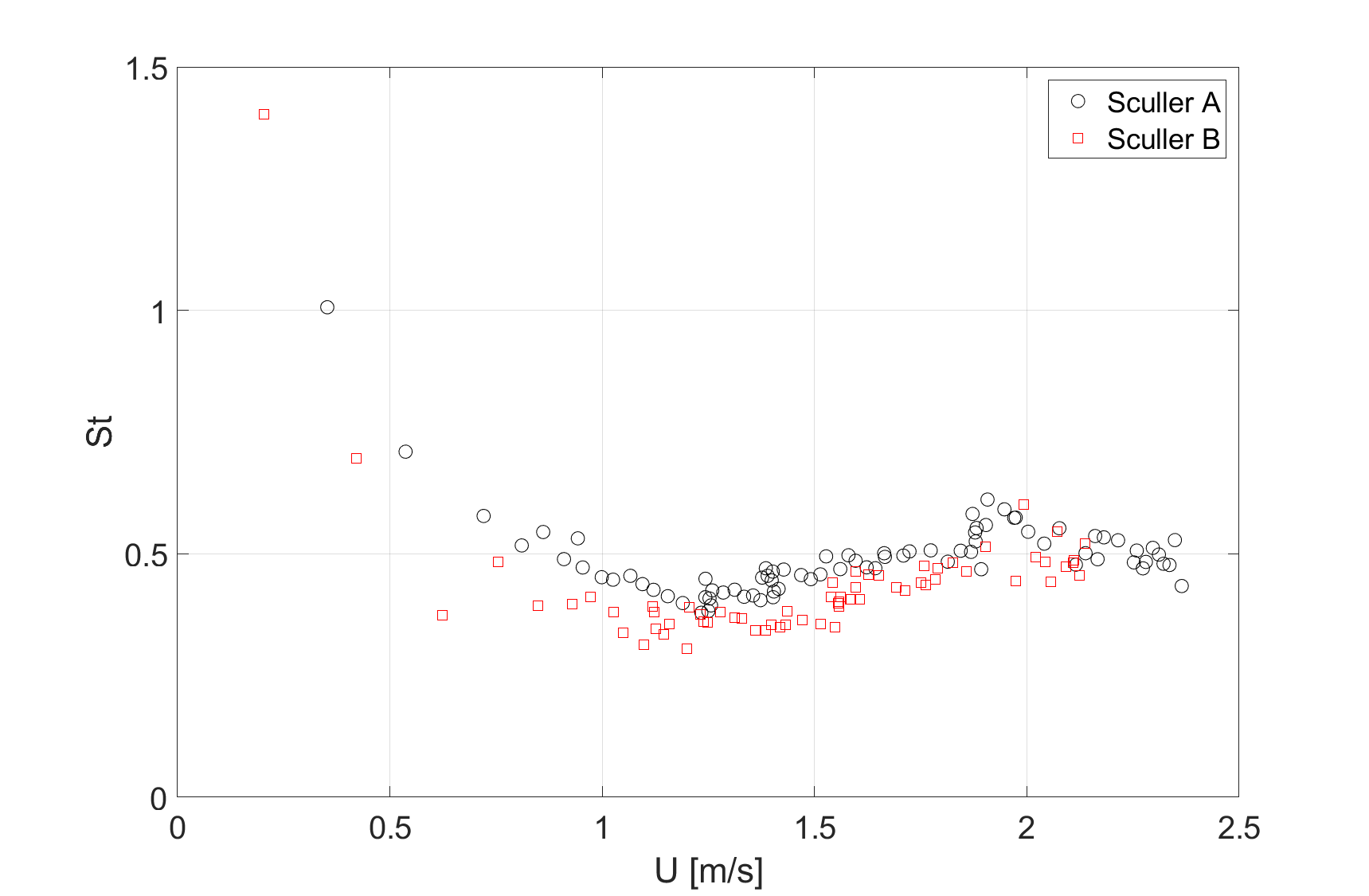}
\caption{Evolution of the Strouhal number (Eq.~\ref{eqn_Strouhal}) for scullers A ($\scalebox{0.7}{\(\bigcirc\)}$) and B (\textcolor{red}{$\square$}) as a function of boat speed $U$. }
\label{Fig_Strouhal}
\end{center}
\end{figure}

Figure~\ref{Fig_Strouhal} shows the evolution of this Strouhal number with the boat velocity. Except at very low velocities, the Strouhal number, $St$, remains approximately $0.5$ in our experiments, indicating that the blade's lateral velocity, that corresponds to $2u_B$, is of the same order of magnitude as that of the boat speed.
This Strouhal number is significantly higher than the values reported in the literature for birds, fish or oscillating foil which typically fall between 0.2 and 0.4 \cite{taylor2003flying, Floryan_2017, katz1978hydrodynamic, read2003forces, labasse2023thrust, seo2026scaling}. However, the situation here is different. In those cases, when wings or fins are on average aligned with the flow, a low Strouhal number corresponds to a low angle of attack of the flow over the constant leading edge. In contrast, for a sculling oar, which is on average perpendicular to the boat's direction of motion, a higher Strouhal number is required to achieve a low angle of attack on the blade's edges.

In order to evaluate the power efficiency of stern-sculling we now compute the instantaneous power injected by the sculler hands into the system.
This human power is given by the scalar product of the hand force $\bm F_M$ by the hand velocity $\bm V_{M}$:
\begin{equation}
P_M(t)=\bm F_M \cdot \bm V_{M}.
\label{eqn_P_M}
\end{equation}

This human power is calculated at all time using the measured forces and kinematics of the oar. We can also compute the instantaneous hydrodynamic propulsive power. The component of $\bm F_H$ in the direction of displacement $x'$ gives the instantaneous propulsive force, and we obtain the hydrodynamic power by multiplying by the boat's speed: 

\begin{equation}
P_H(t)=F_{Hx'}U.
\label{eqn_P_H}
\end{equation}

\begin{figure}[htb]
\begin{center}
\includegraphics[width=0.7\textwidth]{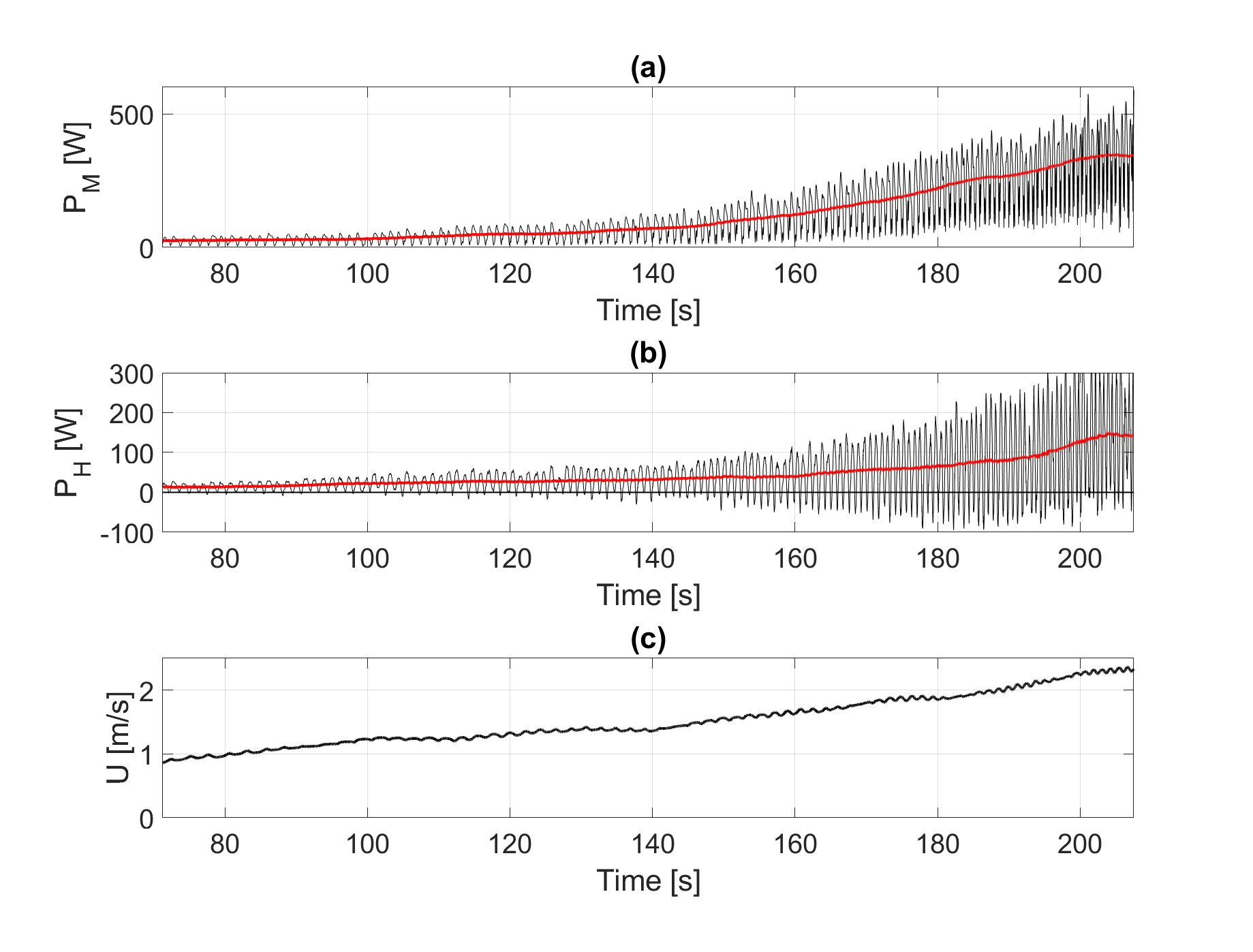}
\caption{Temporal evolution of: (a) the power $P_M(t)$ (Eq.~\ref{eqn_P_M}) produced by the sculler $A$; (b) the corresponding hydrodynamic power $P_H(t)$ (Eq.~\ref{eqn_P_H}); (c) the boat speed $U$ during the run at increasing velocity from $U=0.8\,\text{m/s}$ to $U=2.4\,\text{m/s}$. Red lines represent a sliding average over 500 readings ($10\,\text{m/s}$).} 
\label{Fig_Puissances_instantanees}
\end{center}
\end{figure}

Figure~\ref{Fig_Puissances_instantanees} shows the temporal evolution of $P_M$ and $P_H$ during a run at slowly increasing velocity. Both oscillate twice per period of the oar motion. At the highest velocity, the instantaneous injected power reaches $500\,\text{W}$.

\begin{figure}[htb]
\begin{center}
\includegraphics[width=0.6\textwidth]{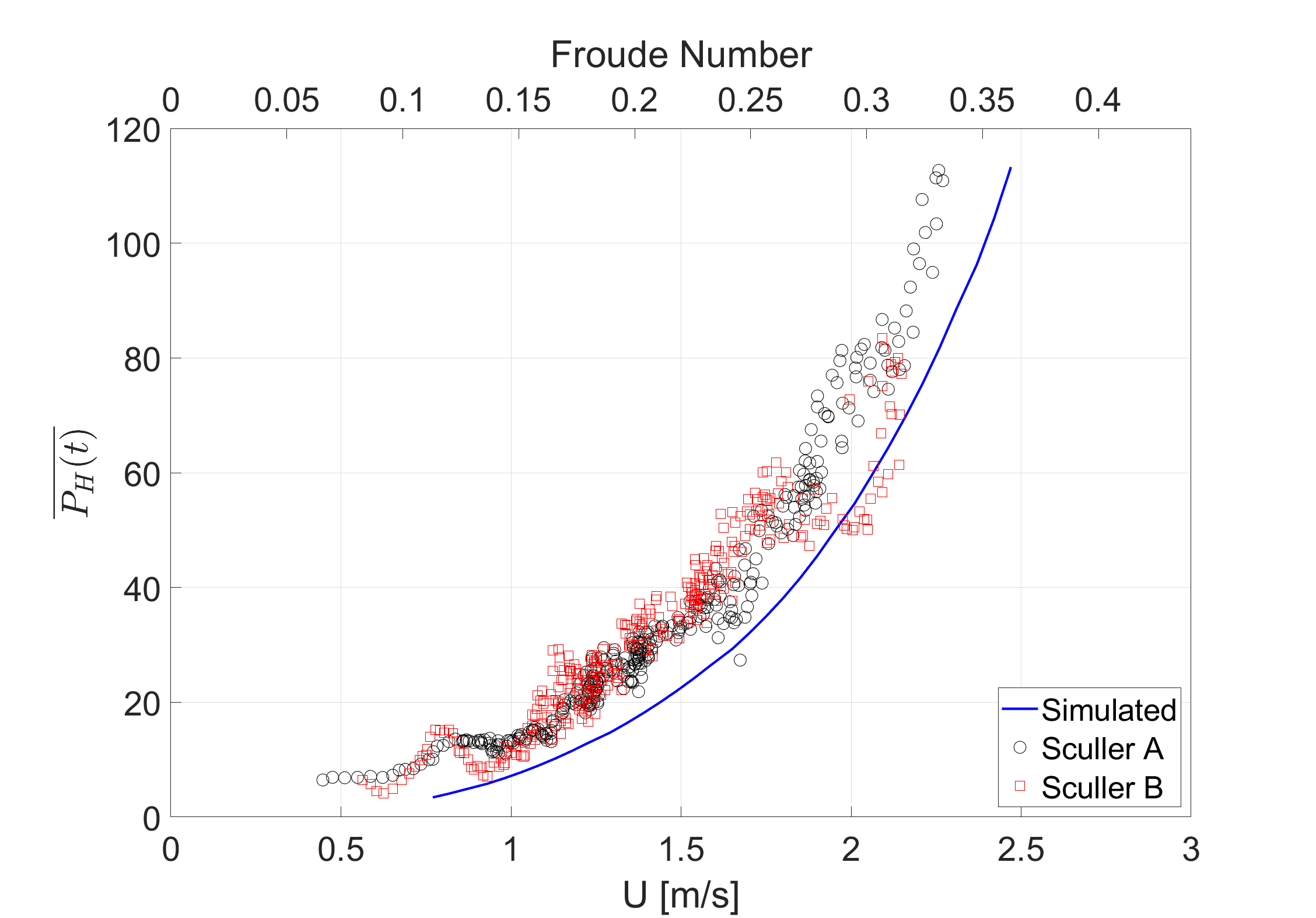}
\caption{Evolution of the mean propulsive power on the boat (averaged over $10$ s.), $\overline{P_H(t)}$, measured for sculler A ($\scalebox{0.7}{\(\bigcirc\)}$) and sculler B (\textcolor{red}{$\square$}), as a function of boat velocity (bottom axis) or Froude number (upper axis). The solid blue line represents a simulation of the hull drag [23]. }
\label{Fig_Trainee-Coque}
\end{center}
\end{figure}

By averaging these powers over a few periods, we define the mean powers. Figure~\ref{Fig_Trainee-Coque} presents the evolution of the mean hydrodynamic power $\overline{P_H(t)}$ generated by the two scullers as a function of the boat speed. In principle, the two curves should coincide, as they both represent the dependence of the hull drag with boat velocity. The slight differences observed may be explained by measurement uncertainties, or by small differences in the hull’s rolling motion, or by differences in wind intensity at a given moment during the two tests. 

A strong increase of $\overline{P_H(t)}$ is observed at high velocity, as is typical when a boat approaches its critical hull speed. To confirmed this, we compute the hull Froude number:

\begin{equation}
 F_r = \frac{U}{\sqrt{gL}}.
\label{eqn_Froude}
\end{equation}

This number compares the boat velocity to the velocity of the gravity waves with a wavelength equal to that of the hull length $L$. When the Froude number increases, the wave drag is known to increase rapidly with a maximum at the critical hull speed when $F_r$ is approximately 0.4~\cite{Rabaud_2014_Ocean}. The Froude values are indicated on the top axes of Figure~\ref{Fig_Trainee-Coque}. In the present tests, the Froude number ranges from 0.07 to 0.37. In that range the propulsive power increases by a factor 10, and at the higher speeds, the hull approaches its critical speed, indicating that wave drag becomes the dominant source of friction. This phenomenon explains why the scullers were unable to go faster, despite producing approximately $300\,\text{W}$.
In Figure~\ref{Fig_Trainee-Coque}, the experimental data on the propulsive power are compared with a computed hull drag obtained with a surface modeling program for ship design \footnote{Free!ship, open source software (\url{https://sourceforge.net/projects/freeship/}) using the 'KAPER' drag calculation method developed by John Winters (\url{http://pointus.varois.free.fr/wa_files/ManualFR_20FreeShip_compressed.pdf}}.
The trends are similar, although the simulation slightly underestimates the experimental drag.

Small oscillations in the mean propulsive power are visible in Figure~\ref{Fig_Trainee-Coque} when velocity increases. These oscillations are reminiscent of hull drag oscillations observed when the bow wave is a sub-multiple of the boat waterline length \cite{Chapman_1972, baar1988developments}. However, in this case, they are not observed for the same Froude number for both scullers. They probably originate from the fact that the boat trajectories are not perfectly straight and that a part of the human power is used to change course.

The power efficiency of stern-sculling propulsion can be defined as the ratio of the mean hydrodynamic propulsive power generated by the oar to the mean human power transferred to the oar:

\begin{equation}
 \eta = \frac{\overline{P_H(t)}}{\overline{P_M(t)}}.
\label{eqn_efficiency}
\end{equation}

\begin{figure}[htb]
\begin{center}
\includegraphics[width=0.6\textwidth]{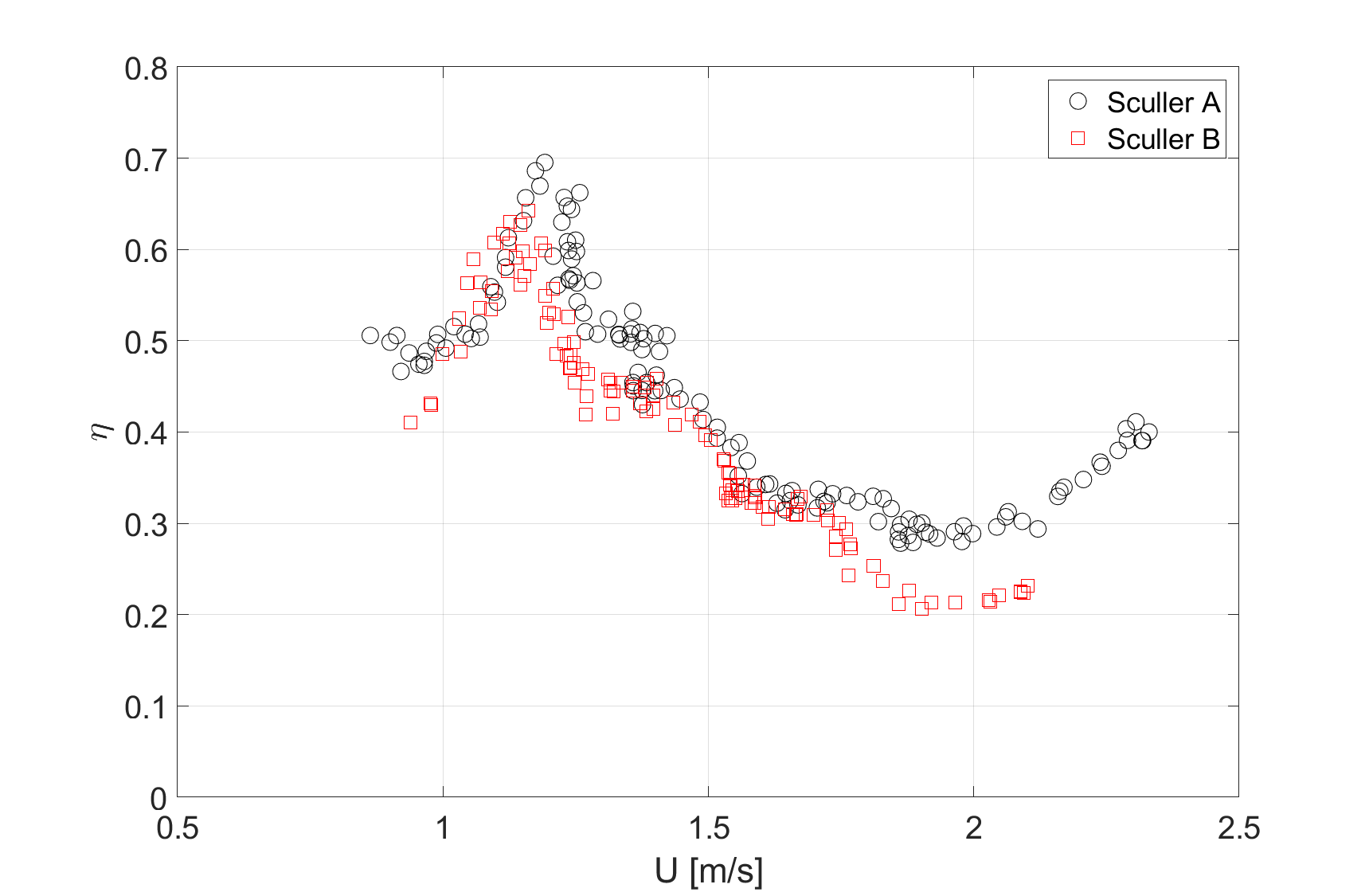}
\caption{Evolution of the power efficiency $\eta$ (Eq.~\ref{eqn_efficiency}) as a function of boat speed $U$ for sculler A (\scalebox{0.7}{\(\bigcirc\)}) and sculler B (\textcolor{red}{$\square$}).}
\label{Fig_efficacity_versus_speed}
\end{center}
\end{figure}

Figure~\ref{Fig_efficacity_versus_speed} shows the evolution of this quantity for increasing boat speed. The efficiency increases at low velocity, reaching $65 \pm 5 $\% at approximately $1.2\,\text{m/s}$ and decreases at higher velocity ($25 \pm 5 $\% at approximately $2\,\text{m/s}$). As a point of comparison, the efficiency of a marine propeller is of the order of 75\% for its designed speed but much smaller at other speeds \cite{barnitsas1981kt}. In classical rowing, the best efficiency with two lateral oars is also of the same order of 75\% \cite{cabrera2006propulsive}. 
In the case of the oscillating propeller developed for kayaks and described in \cite{augier2015experimental}, efficiency increases linearly with the frequency of movement, up to 60\%. This trend is enabled by the deformation of the fin. 

The low performance observed at low speeds may be due to suboptimal stroke techniques, which are rarely practiced. At highest speeds---such as when a propeller operates above its nominal speed---efficiency is likely to decrease because, for a fixed roll angle $\delta$, the hand velocity cannot be increased proportionally, resulting in an angle of attack of the flow on the blade that becomes too small.

\section{Analysis of the flow and the forces acting on the blade}

In the previous sections, we analyzed the oar's kinematics and measured the forces and the propulsion efficiency of stern-sculling. In this section, our aim is to describe the flow over the blade in order to gain a better understanding of the origin of the propulsive forces during the various phases of the oar motion.

\subsection{Incident flow on the blade}

\begin{figure}[htb]
\begin{center}
\includegraphics[width=0.9\textwidth]{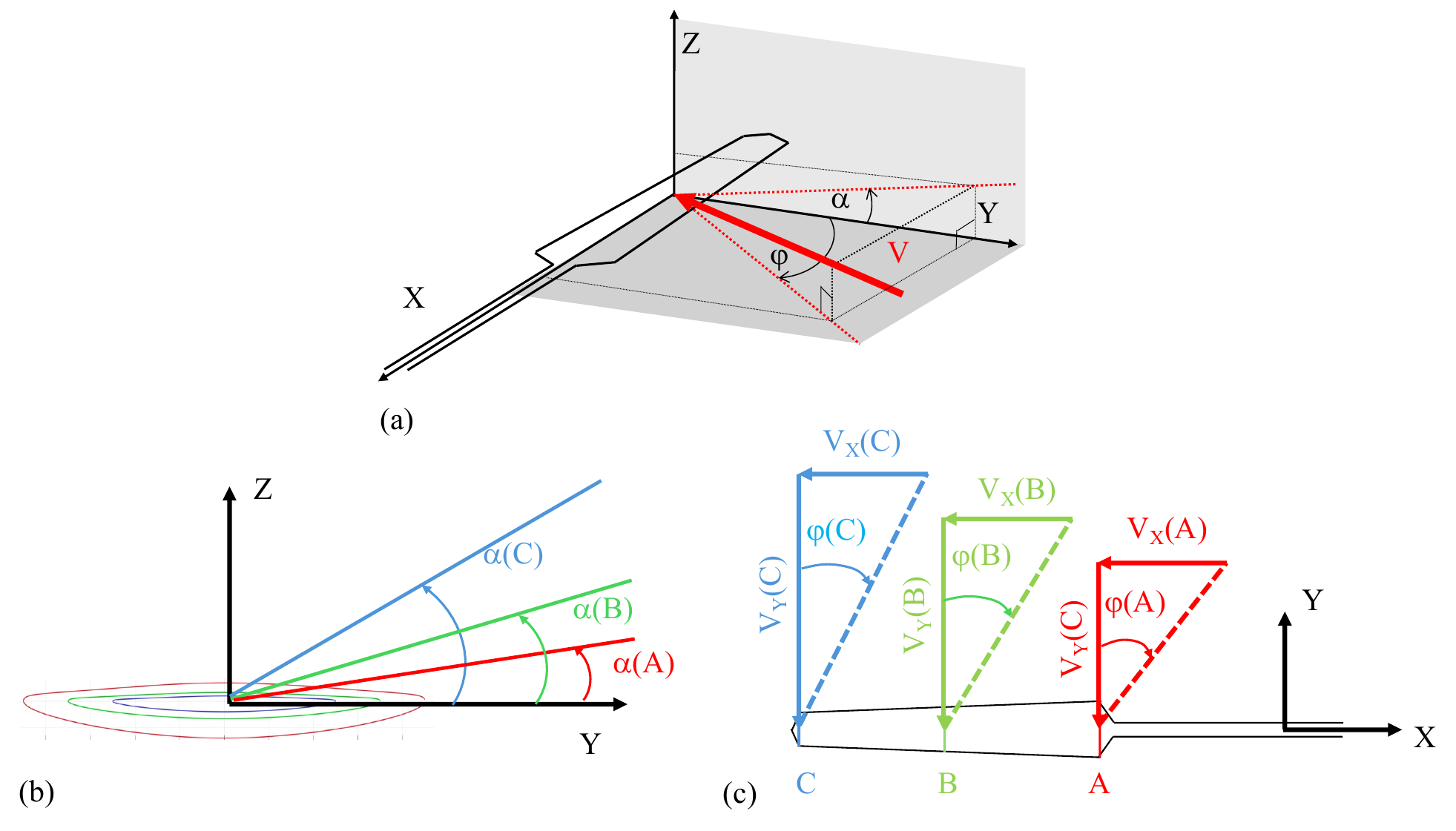}
\caption{Incident flow angles in the blade's frame, represented during phase 1: (a) definitions of angles $\alpha(X,t)$ and $\varphi(X,t)$ of the incident flow; (b) evolution of $\alpha$ in sections $A,B$ and $C$ of the blade; (c) evolution of $\varphi$ in the same sections. The angle $\alpha$ increases from A to C, while $\varphi$ decreases. For example, at time $t=109.4\,\text{s}$  in Figure~\ref{3_Evolutions}, $U=1.2\,\text{m/s}$ and $\gamma = 0^\circ$, $\beta = 31^\circ$, $\delta = 38^\circ$ and angular velocity $\dot \gamma = -48^\circ/\text{s}$. At this instant: $\alpha(A) = -3 ^\circ$, $\alpha(B) = 3 ^\circ$ and $\alpha(C) = 8 ^\circ$; $\varphi(A)=42 ^\circ$, $\varphi(B)=35 ^\circ$, $\varphi(C)=30 ^\circ$; $V_Y(A)=1.25$ ~\text{m/s}; $V_Y(B)=1.4\,\text{m/s}$ and $V_Y(C) = 1.7\,\text{m/s}$.}
\label{Fig_Flow_Oar}
\end{center}
\end{figure}

 The incident velocity of the water with respect to the blade can be calculated at each point $(X,0,0)$ on the axis of the blade. This incident velocity denoted $\bm V(X,t)=(V_X, V_Y,V_Z)$ in the oar frame. It is the opposite of the vector sum of boat speed $\bm U$ and of the velocity of point $(X,0,0)$ with respect to the mean boat frame.
At each instant, this incident flow varies in direction and magnitude along the length of the blade.
Two angles are necessary to characterize the local orientation of the incident water flow (Fig.~\ref{Fig_Flow_Oar}a):

\begin{itemize}
 \item First, an angle of attack $\alpha$ defined by:
\begin{equation}
 \tan \alpha(X,t) = \frac{V_Z}{V_Y}.
\label{eqn_alpha}
\end{equation}
Since the hydrodynamic force corresponds to propulsion, the water typically reaches the upper face of the blade, so $V_Z < 0$. This angle $\alpha$ corresponds to the standard angle of attack on a wing if the flow was in the transverse $(Y,Z)$ plane (phase 1) and coming from $Y> 0$. The angle of attack is $\pi - \alpha$ when the flow comes from $Y<0$ (phase 3).

 \item Second, a sweep angle, $\varphi$, defined by:
\begin{equation}
 \tan \varphi(X,t)= \frac{V_X}{V_Y}.
\label{eqn_varphi}
\end{equation}

 It is a measure of the component along the blade (spanwise) of the incident flow. It is used, for example, for aircraft with swept-back wings. 
\end{itemize}

Due to the rotation of the oar around $D$, the blade velocity increases with $X$ and both these angles evolve with $X$. The angle of attack $\alpha$ in phase 1 increases progressively toward the bottom of the blade (Fig.~\ref{Fig_Flow_Oar}b). Note that this angle is slightly negative at the top of the blade in phase 1 (and greater than $180 ^\circ$ in phase 3). This is similar to what occurs on the blade of a wind-turbine or on a propeller blade, this is why their are usually twisted. However, this is not possible here because the leading edge alternates over time. Furthermore, since the oar has an inclination $\beta$, the water flow in the blade's frame has a component in the $-X$ direction at the origin of the sweep angle of the incident flow. The sweep angle of the flow $\varphi(X)$ decreases progressively toward the bottom of the blade in phase 1 (Fig.~\ref{Fig_Flow_Oar}c).

\begin{figure}[htb]
\begin{center}
\includegraphics[width=0.95\textwidth]{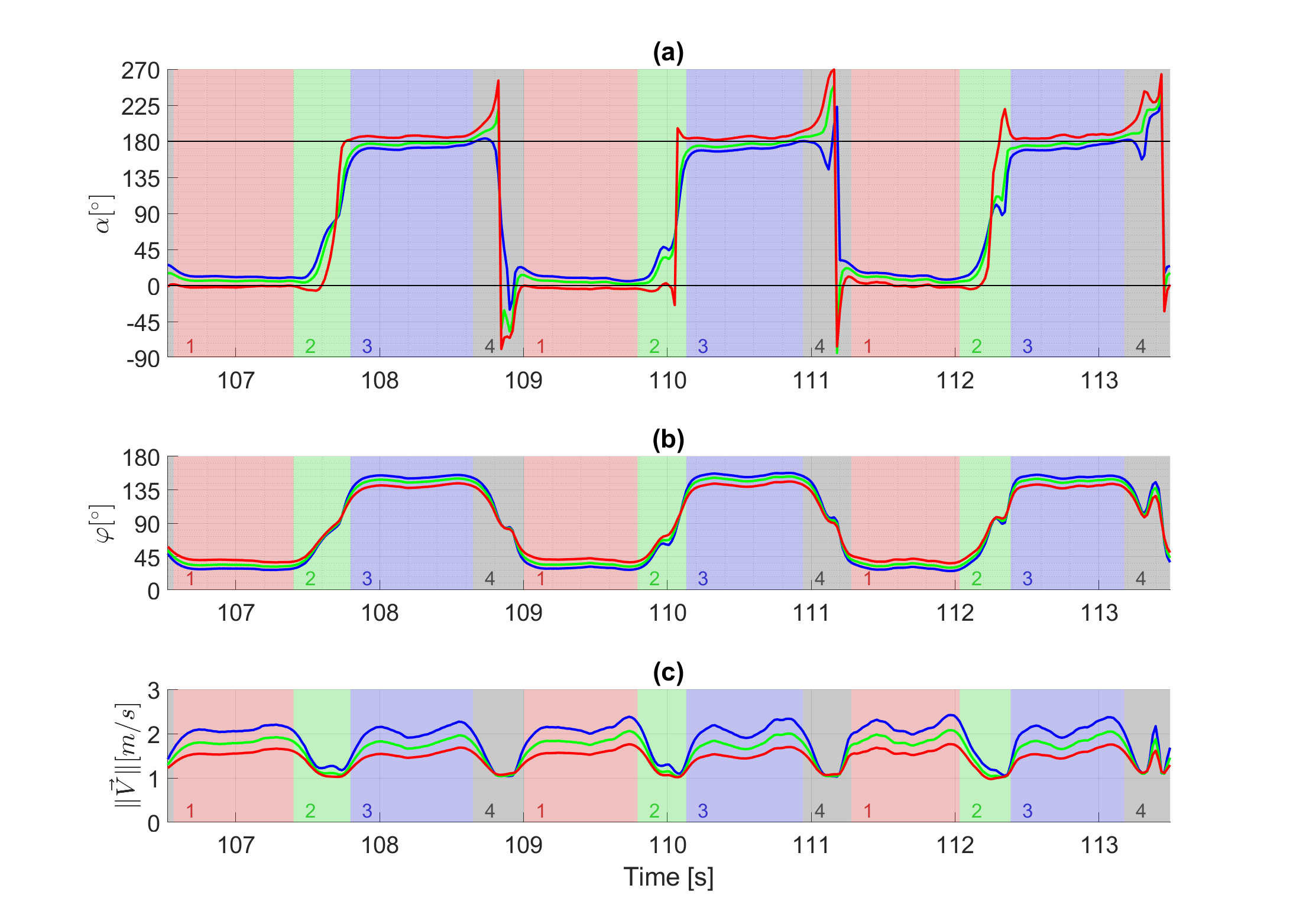}

\caption{Temporal evolution during three cycles, for a run at $U = 1.2$ ~\text{m/s}, of the incident water flow in the three blade sections $A$, $B$ and $C$ as defined in Figure~\ref{oar_drawing}: top ($A$, $X_A= - 1.26\,\text{m}$, red), middle ($B$, $X_B= - 1.73\,\text{m}$, green) and bottom ($X_C= - 2.20\,\text{m}$, blue). Background colors indicate the four phases of the oar motion: (a) 2---D incident angle $\alpha(X,t)$; (b) sweep angle $\varphi(X,t)$; (c) modulus of the incident velocity vector $\|\bm V(X,t) \|$.}
\label{Fig_angles_Incidence}
\end{center}
\end{figure}

During the periodic motion of the blade, these angles evolve in time (Fig.~\ref{Fig_angles_Incidence}), crossing $90 ^\circ$ during phases 2 and 4, when the lateral velocity of the blade vanishes and $\delta = 0 ^\circ.$
Figure~\ref{Fig_angles_Incidence}c shows that the modulus of the incident velocity increases with the distance from the oarlock.

\subsection{Estimation of the hydrodynamic force during lateral strokes}

During phases 1 and 3, where large lateral movements of the oar take place, the roll angle varies slowly and the angle of attack, $\alpha$ or $\pi -\alpha$, remains small everywhere on the blade (Fig.~\ref{Fig_angles_Incidence}), indicating that the water flow remains essentially attached and that lift is the primary generated force. 
With a blade width of approximately $150\,\text{mm}$ and a water velocity consistently exceeding $1\,\text{m/s}$, the advection time of the flow on the blade is shorter than the duration of the various phases of the oar’s movement, allowing the use of a quasi-static approximation. As the incident flow on the blade changes its angles with $X$, the lift and drag directions also evolve with $X$. Thus, we estimate them on a set of 10 narrow transverse slices of the blade, assuming that in each slice the flow is transverse (neglecting the spanwise flow) and with an angle of attack $\alpha$ that depends on the slice, inspired by aerodynamic strip theory \cite{dowell1999modern}.
With this approach, no hydrodynamic force will be found along $X$, in agreement with initial findings.

\begin{figure}[tbh]
\begin{center}
\includegraphics[width=0.6\textwidth]{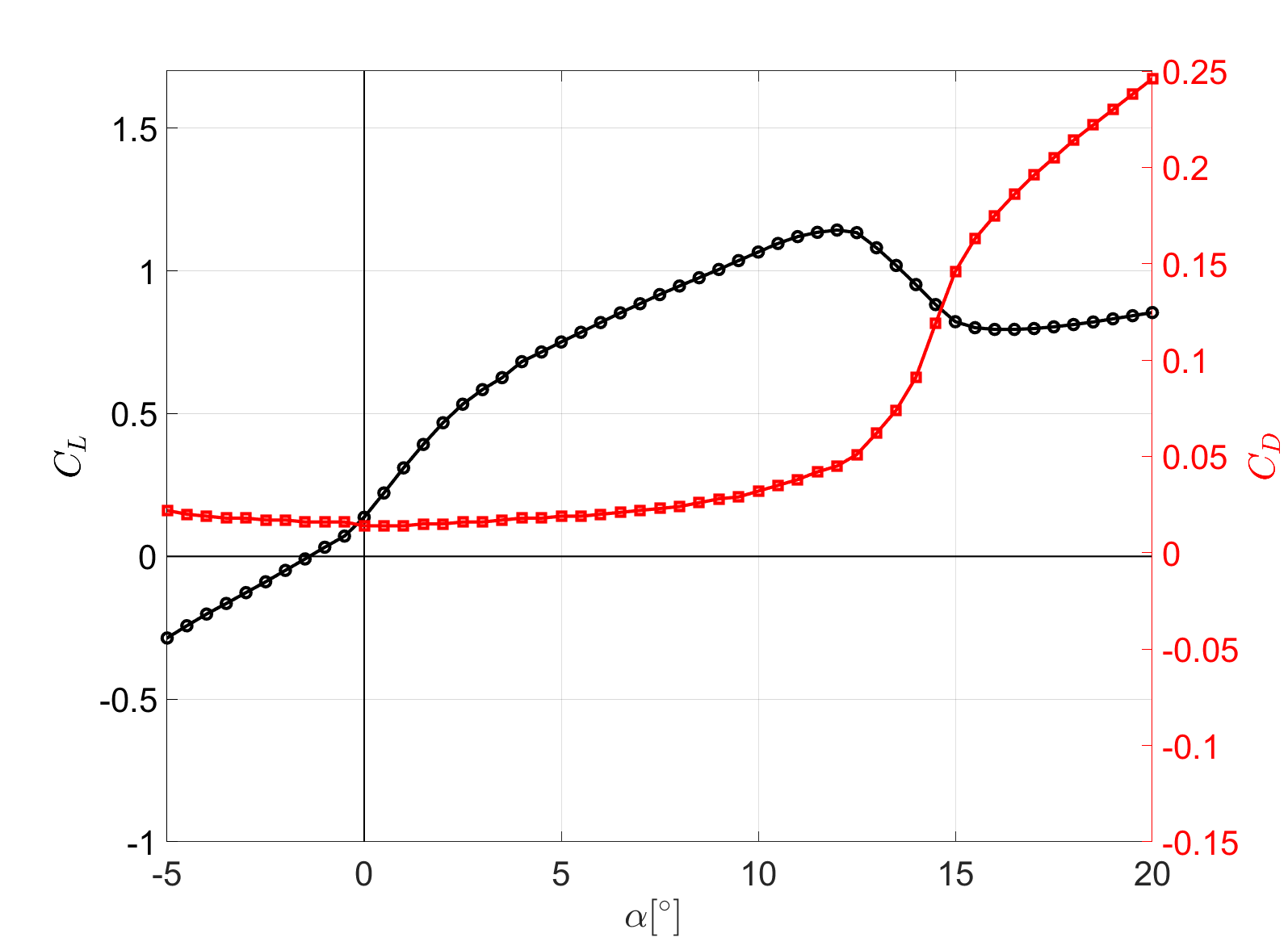}
\caption{Lift ($C_L$, $\bullet$, left scale) and drag ($C_D$, \textcolor{red}{$\rule{1ex}{1ex}$}, right scale) coefficients simulated with XFOIL for the 2-D profile corresponding to Fig.~\ref{oar_drawing}b, as a function of the angle of attack $\alpha$ at $\mathrm{Re}= 200\,000$.}
\label{CD_CL_simulated}
\end{center}
\end{figure}

As the flow is attached, or only partially separated, we compute for each slice the lift and drag coefficients $C_L$ and $C_D$ using the open source program XFOIL’s viscous flow solver, which couples a panel method for the inviscid potential flow with a boundary-layer model that accounts for viscous effects and laminar-to-turbulent transition. This computationally efficient approach enables the prediction of pressure distributions and hydrodynamic coefficients over a range of Reynolds numbers and angles of attack and has been widely validated against CFD and experimental data (\cite{drela1989,morgado2016,gunel2016comparison}). In the present study, XFOIL calculations are performed assuming an infinite 2-D wing with the experimental profile of the blade (fig.~\ref{oar_drawing}b) and a Reynolds number $Re = 200\,000$. This Reynolds number corresponds, at the middle of the blade, to an incident velocity of 1.4 ~\text{m/s}, which is a good order of magnitude during phase 1 or 3. These 2-D lift and drag coefficients are computed as a function of the angle of attack $\alpha$, and the results are presented in Figure~\ref{CD_CL_simulated}.

As the blade profile presents a small camber, lift is positive for $\alpha=0 ^\circ$ ($C_L=0.14$) and increases progressively until $\alpha=12 ^\circ$, where stall begins. The drag coefficient $C_D$ increases slowly until $\alpha=12 ^\circ$ and more rapidly when $C_L$ starts to decrease. Stall occurs relatively gradually, likely because the leading and trailing edges are rounded.

\begin{figure}[htb]
\begin{center}
\includegraphics[width=0.6\textwidth]{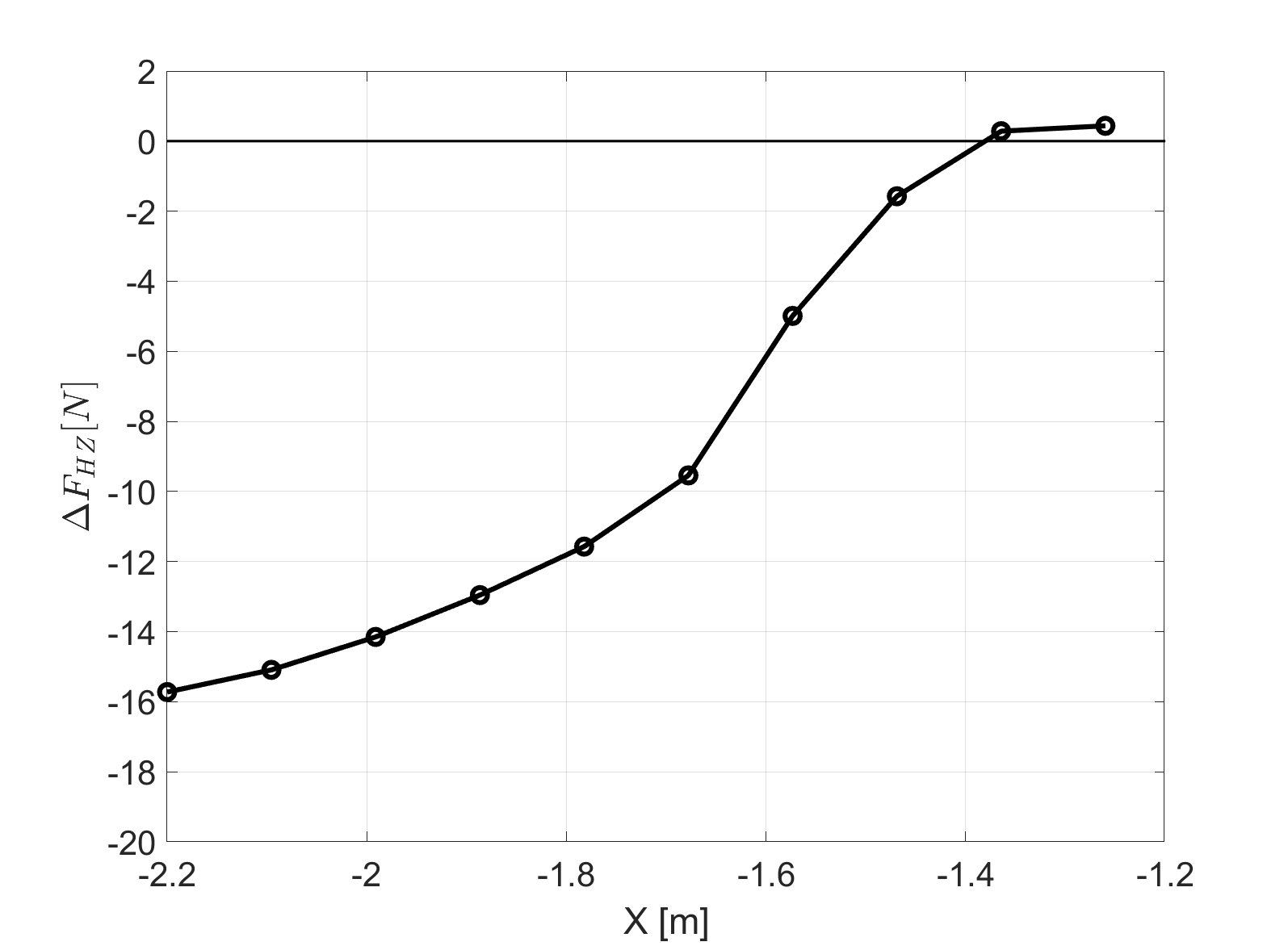}
\caption{Numerical results for the component $\Delta F_{HZ}$ of the hydrodynamic force for the 10 selected slices of the blade at $t=109.4\,\text{s}$ (see Figure \ref{3_Evolutions}).}
\label{Portance_theorique}
\end{center}
\end{figure}

Since the lift-to-drag ratio is large and as we are interested in estimating the propulsive force of the oar in the lateral strokes, we now focus only on the lift on each of the slices of the oar.
 We select a particular instant when the yaw angle $\gamma$ is closest to zero, corresponding to the middle of phase 3 when $\alpha$ and $\|\bm{V} \|$ are relatively stable ($t=109.4$ s in Fig.~\ref{Fig_angles_Incidence}). 
 Using the experimental data at this instant, we determine the mean angle of attack on 10 slices. Given the lift surface $\Delta S$ of the slice and the water density $\rho_w$, we compute the lift, $\Delta \bm F_L$, acting on this slice: 
 \begin{equation}
 \Delta \bm F_L = \frac{1}{2}\rho_w (V_Y^2+V_Z^2)C_L \Delta S.
\label{eqn_lift_slice}
\end{equation}
 By projecting the lift force into the oar frame, we obtain the components $\Delta F_{HZ}$ of the lift force for each slices. Figure \ref{Portance_theorique} presents these 10 contributions at this instant. The results show that the hydrodynamic force is larger at the bottom of the blade because the angle of attack and the incident velocity are larger, and that $\Delta F_{HZ}$ is slightly positive (non-propulsive) near the surface as $\alpha(A) = -3 ^\circ$.

Summing the contributions of all the slices, we obtain the total component $F_{HZ}$. At this instant $t$, our calculation yields $F_{HZ} = - 85\,\text{N}$, while the experimental measurement is $- 77\,\text{N}$. Projecting these results in the direction of the boat's mean displacement gives the instantaneous force $F_{Hx'} = 35\,\text{N}$, compared to the experimental value of $31\,\text{N}$ at this instant. These results are reasonable, considering that we neglected drag force on each slice, which has a small contribution to the propulsion component. Accounting for this drag is not straightforward and would require many approximations ({\textit{e.g.}}, a global induced drag correction with an unknown Oswald coefficient \cite{anderson2011ebook}).

\subsection{Estimation of the hydrodynamic force during backward strokes}

In the same way, we will now estimate the hydrodynamic force acting during the backward strokes. In these phases 2 and 4 the angles of attack are large (Figure~\ref{Fig_angles_Incidence}a) suggesting flow separation and consequently an hydrodynamic force resulting primary from drag.
As for the lift evaluation, we could in principle compute the small drag force on each blade slices using the relation: 
\begin{equation}
 \Delta F_D= \frac{1}{2}\rho_w (V_Y^2+V_Z^2)C_D\Delta S.
\label{eqn_drag_slice}
\end{equation}
However, even using a large value of $C_D$, for example $C_D=1.5$ \cite{hemon2018hydrodynamic}, and summing over all the slices, we obtain simulated $F_{HZ}$ ranging from $-2\,\text{N}$ to $-10\,\text{N}$ during phase 2. These values are significantly underestimated compared to the experimental $F_{HZ}$, which ranges from $-20\,\text{N}$ to $-45\,\text{N}$ in this phase. This discrepancy arises because during these phases, the blade acceleration is significant. There is therefore a substantial added mass term that corresponds to the change in the kinetic energy of the fluid around the blade.
This force is generally expressed as:

\begin{equation}
 \bm F_{am}=-\kappa \rho_w \mathcal{V} \frac{d\bm V(t)}{dt},
\label{eqn_addedmass}
\end{equation}

 where $\mathcal{V}$ is the object volume and $\kappa = 1$ for a cylinder. Since the angle of attack on the blade is close to $\pi/2$,  we can estimate the volume here to be that of the cylinder containing the blade, we find $\mathcal{V}= 0.015\,\text{m}^3$. With an acceleration of $4.5\,\text{m/s}^2$ at $t= 110\,\text{s}$ of Figure \ref{3_Evolutions}, we obtain $F_{am}=-70\,\text{N}$. This shows that the added mass force dominates the steady-state drag force and could well explain the measured forces in these phases, even though a time-resolved comparison is beyond the scope of this paper.

\section{Discussion and conclusion}
\label{sec:conclusion}

Stern-sculling is a periodic propulsion technique characterized by the '$\infty$'-shaped movement of an oar located at the stern of the boat. This periodic motion can be broken down into four phases: two lateral strokes and 2 backward strokes (rotation phases at the extremities of the "$\infty$" symbol, during which the blade is temporarily perpendicular to the flow). By analyzing the forces and kinematics during the motion, we have shown that, in the lateral phases, propulsion is mainly due to lift forces on the blade, whereas in the backward phases, propulsion results from drag forces induced by the blade's acceleration. The propulsive force oscillates significantly over time, reaching minima close to zero twice per oar cycle.

This traditional propulsion technique has proven to be highly efficient at low speeds but far less so at higher speed: its efficiency decreases from 65\% to 25\% as boat speed increases in the studied setup. We calculated the angles of attack of the water flow at three sections of the blade and showed that, during the lateral motion, this angle is slightly negative at the top of the blade, which has a small negative effect on propulsion, and strongly positive at the bottom of the blade, suggesting a detachment of the water streamlines. The decreases in efficiency at high boat speed was then attributed to two factors: (i) in the lateral phases, a favorable angle of attack of the flow on the blade requires a high lateral velocity, which becomes bio-mechanically difficult to achieve; (ii) in the backward phases, it becomes harder to move the oar backward fast enough to maintain a propulsive drag on the blade and to prevent the oar to leave the notch.

Several strategies exist---though not studied in this paper---to overcome these limitations at high boat speeds: (a) the sculler can set the oar more horizontal by decreasing the oar pitch angle {$\beta$}, although this reduces the projection of hydrodynamic forces along the propulsive axis; (b) the sculler can increase the oar's length, with the same blade, to limit the evolution of the angles of attack along the blade. However, this would result in large-amplitude hand movements if the upper part of the oar is increased proportionally, or excessively large forces for the sculler if the length $DM$ is not increased; c) the sculler can use a downward-curved blade to decrease the incident angle at the tip of the blade, which also improves the projection of the hydrodynamic force. The drawback of this solution is the difficulty involved in manufacturing and handling the oars.

In the present work, the maximum speed is directly linked to the hull's critical speed, as evidenced by the Froude number ($Fr = 0.35$) at the highest velocities. To mitigate the rapid increase in wave drag, a longer hull could be used; however, the system would still experience the efficiency loss at high speeds previously described. The maximum speed achievable by stern-sculling therefore depends on a combination of the hull and oar properties and the associated kinematics. In practice, scullers adjust the length of their oars to suit the boat and its intended use, whether for endurance, speed or maneuverability.

The present study was carried out with a relatively rigid oar to simplify kinematic and geometric reconstruction. 
In practice, scullers may seek a degree of flexibility in the $X$, $Y$ or $Z$ directions, particularly to smooth out the forces and to provide a degree of tolerance in the kinematics. Flexibility may also be incorporated into the oar’s design to improve efficiency, particularly by reducing the angle of attack at the tip of the blade, thereby delaying stall. For example, flexibility in the streamwise direction can reduce the camber (a strategy exploited in nature or in micro air vehicles to delay stall \cite{tiomkin2021review,ramananarivo2011rather}) or induce spanwise twist (a concept studied for marine propulsion \cite{YoungPropFSI} or oscillating blades \cite{jin2026spanwise}). This point would be worth exploring in more detail in the future, keeping in mind that designing the ideal oar for a skiff involves many trade-offs, including the power generated by the sculler.

\begin{acknowledgments}

We thank M. Benoît, the students of "Navigodille" project from IMT Atlantique School, O. Le Rest, O. Guillerm, S. Guerin, M. Pitel and to association TYAM from Lampaul-Plouarzel who lent us the boat. M.R. thanks I. Delbende, Th. Faure, P.-Y. Lagrée, F. Moisy and M. Rossi for stimulating discussions.

\end{acknowledgments}

\bibliographystyle{unsrt}
\bibliography{Biblio_Sculling}

\end{document}